\documentclass[preprint,amsmath,amssymb,aps]{revtex4-2}

\usepackage{graphicx}
\usepackage{dcolumn}
\usepackage{bm}
\usepackage{bigints}

\newcommand*\pFq[2]{{}_{#1}F_{#2}}

\begin{document}


\title{First-passage time statistics for non-linear diffusion}

\author{Przemys\l{}aw Che\l{}miniak}
\email{geronimo@amu.edu.pl}
\affiliation{Institute of Spintronics and Quantum Information, Faculty of Physics,
Adam Mickiewicz University, Uniwersytetu Pozna\'nskiego 2, 61-614 Pozna\'n, Poland}

\date{October 2022}

\begin{abstract}
Evaluating the completion time of a random algorithm or a running stochastic process is
a valuable tip not only from a purely theoretical, but also pragmatic point of view. In
the formal sense, this kind of a task is specified in terms of the first-passage time
statistics. Although first-passage properties of diffusive processes, usually modeled
by different types of the linear differential equations, are permanently explored with
unflagging intensity, there still exists noticeable niche in this subject concerning
the study of the non-linear diffusive processes. Therefore, the objective of the present
paper is to fill this gap, at least to some extent. Here, we consider the non-linear
diffusion equation in which a diffusivity is power-law dependent on the
concentration/probability density, and analyse its properties from the viewpoint of the
first-passage time statistics. Depending on the value of the power-law exponent, we
demonstrate the exact and approximate expressions for the survival probability and the
first-passage time distribution along with its asymptotic representation. These results
refer to the freely and harmonically trapped diffusing particle. While in the former case
the mean first-passage time is divergent, even though the first-passage time distribution
is normalized to unity, it is finite in the latter. To support this result, we derive the
exact formula for the mean first-passage time to the target prescribed in the minimum of
the harmonic potential.
\end{abstract}

\maketitle

\newpage
\section{\label{sec_1}Introduction}

The first-passage time statistics has attracted much attention of the scientific community
for more than the past century~\cite{Schr1915,Smol1915,Kram1940} and last until today.
Noteworthy is the collection of excellent books~\cite{Redn2001,Osha2014} and review
papers~\cite{Bray2013,Beni2014,Iyer2016,Greb2020} that have been published in recent years.
The research topic included in these publications reveals that the first-passage phenomena
appears in such diverse disciplines as applied mathematics, physics, chemistry, biology and
even economics and finance, to name but a few examples. The fundamental concept used in
exploration of this kind of phenomena is the first-passage time distribution, according to
which the very time is thought of as a random variable. Its average value, in formal terms
the first moment of the time distribution, is called the mean first-passage time, also known
as hitting time, crossing time or exit time, depending on the specific problem. For this
reason, the dynamics of systems studied as part of the above-mentioned disciplines are
typically represented by stochastic processes~\cite{Gard2004,Kamp2007}. The mean first-passage
time is then the first moment when the stochastic process reaches a predetermined state,
starting form some initial state. The illustrative example is the mean time for a Brownian
particle to hit a prescribed spatial position; the mean time for an enzyme to recognize and
interact with a substrate molecule; and the mean time when the stock prise of a product exceeds
a certain threshold.

A well-known prototype of the stochastic process is a diffusive motion, whose first-passage
properties are the main objective of the present paper. In recent decades, the first-passage
time statistics has been explored for a variety of diffusive processes such as ordinary
diffusion~\cite{Mahn2009}, diffusion in external potentials~\cite{Greb2015}, in an Euclidean
domains~\cite{Holc2014,Greb2016,Greb2019}, hierarchical or fractal-like porous media
~\cite{Cond2007,Nguy2010} and heterogeneous media~\cite{Vacc2015}. Of particular interest were
also continuous-time random walk~\cite{Greb2018}, fractional Brownian motion~\cite{Molc1999},
L{\'e}vy flights and walks~\cite{Pada2019,Paly2019} and self-similarity of diffusions' first
passage times~\cite{Elia2021}. For a dozen years the current research topic, {\it inter
alia} in the context of the first-passage problems, are diffusive processes intermittent by
stochastic resetting~\cite{Evan2011,Evan2020}. The joint feature of the most listed processes
is their space-time dynamics which are formalized in terms of more or less elaborate
{\it linear} partial differential equations. This fact raises the legitimate question about the
first-passage properties of the {\it non-linear} diffusive processes.  In this paper, we focus
on the special variant of the non-linear diffusion equation, known as {\it porous medium
equation}~\cite{Vazq2007}, in which a diffusion coefficient is power-law dependent on the
probability density or concentration of particles. In addition, this relation is parameterized
by the power-law exponent that will be assumed to be positive and constant. The porous medium
equation has found many applications in the study of such disparate transport phenomena as
compressible gas flow through porous media~\cite{Bare1990}, heat propagation occurring in
plasma~\cite{Berr1978}, groundwater flow in fluid mechanics~\cite{Koch1948}, population
migration in biological environment ~\cite{Gurt1977,Murr2002}, the diffusion of grains in
granular matter~\cite{Chri2012} and gravity-driven fluid flow in layered porous
media~\cite{Prit2001}.

Here, we study the porous medium equation form yet another perspective, namely the
first-passage time statistics. Although such a problem has already been analysed for the
fractional non-linear diffusion equation in Ref.~\cite{Wang2008}, some of the results
presented in that paper seem to be at least controversial. The reason for our criticism is the
improper assumption made by the authors of the cited work that the target point, to which the
diffusing object moves and then reaches for the first time, comprises the totally absorbing
well. To uphold our objection, we will evidently show that trying to solve the non-linear
diffusion equation in the presence of the absorbing boundary condition leads to some kind of
contradiction. Instead, we will demonstrate the alternative approach to the first-passage time
problem concerning the non-linear diffusion.

Let us clarify that throughout this paper the diffusive motion will be restricted to the
semi-infinite interval with the target point located at the origin. For such a system we
determine the survival probabilities, depending on the power-law exponent characterizing
the relation between the diffusion coefficient and the probability density. Surprisingly,
this thread has been omitted for some reasons in Ref.~\cite{Wang2008}. In general, the
survival probability defines a likelihood that the first-passage event has not occurred until
a given time interval. It is equal to unity at the initial moment of time and then immediately
begins to diminish in time. However, by contrast with the ordinary diffusion, the time
course of the survival probability for the non-linear diffusion takes place in two phases.
Through the first period of time its value constantly remains equal to unity and only in
the second phase it monotonically decreases to zero. We show that such a progress of the
survival probability in time results from the time evolution of the probability density
whose domain is limited to the support of finite extend, outside of which this function
disappears. Consequently, it takes some time for the front of the probability density to
reach the prescribed target point. Armed with the survival probability, we calculate
the first-passage time distribution defined as the time derivative of that former quantity
with a minus sign. We point out that due to the long-time tails of the first-passage time
distributions obtained for the non-linear diffusion in the semi-infinite interval, their
first moment, namely, the mean time to the origin, is divergent. We argue that this hitting
time becomes finite as far as a diffusive motion occurs in a bounded domain of a space. To
justify our statement, we consider as an example the harmonically trapped particle whose
dynamics are determined by the non-linear diffusion equation.

The structure of the paper is as follows. In the subsequent section we give a brief overview
of the non-linear diffusion equation. Sec.~\ref{sec_3} is reserved to revise the basic
concepts of the first-passage time statistics. Here, we also give the reason for which
the absorbing boundary conditions are not compatible with the non-linear diffusion equation.
In Sec.~\ref{sec_4} we present the main results regarding first-passage properties of the
non-linear diffusion. The analysis of this process in the harmonic potential is performed in
Sec.~\ref{sec_5}. We summarize our results in Sec.~\ref{sec_6}.

\section{\label{sec_2}Non-linear diffusion equation}

In what follows, we restrict our studies of the non-linear diffusion along with its
first-passage time statistics to one dimension. Before we formulate a special type of the
equation describing this process, let us firstly consider its more general form (see for
example~\cite{Debn2012}), namely
\begin{equation}
\frac{\partial}{\partial t}p(x,t)=\frac{\partial}{\partial x}\left(\mathcal{D}
\left[\,p(x,t)\right]\frac{\partial}{\partial x}p(x,t)\right).
\label{eq_1}
\end{equation}
By definition, this is the non-linear partial differential equation for the function $p(x,t)$,
which in a physical sense may stand for, depending on the context, the concentration of
diffusing particles, where $x$ is the distance from some initial position and $t$ is the
time, or the probability density function (PDF) of finding a diffusing particle in the
location $x$ at time $t$. In this paper we will consequently use the latter interpretation.
The reason for the non-linear nature of Eq.~(\ref{eq_1}) is a direct dependence of the
diffusivity $\mathcal{D}[p(x,t)]$ on the PDF through which it also depends on the variables
$x$ and $t$. Therefore, to specify the particular form of Eq.~(\ref{eq_1}) we have yet to
establish a specific relationship between the diffusion coefficient and the PDF. Due to many
interesting and practical applications that have attracted considerable attention within
scientific community~\cite{Fasa1986}, we define this relation by the power-law function
\begin{equation}
\mathcal{D}=\mathcal{D}_{0}\left(\frac{p(x,t)}{p_{0}}\right)^{\sigma}.
\label{eq_2}
\end{equation}
In this expression $p_{0}$ denotes a constant reference value of a probability density, whereas
$\mathcal{D}_{0}$ is the diffusivity at that reference value. The power-law exponent $\sigma$
is a certain parameter. Only in the particular case for $\sigma\!=\!0$, Eq.~(\ref{eq_1})
converts into the linear diffusion equation with a diffusion constant $\mathcal{D}_{0}$.

To give Eq.~(\ref{eq_1}) a more convenient form, we now rewrite the diffusion coefficient
in Eq.~(\ref{eq_2}) so that $\mathcal{D}=Dp^{\sigma}(x,t)$. Here, the parameter
$D=\mathcal{D}_{0}/p^{\sigma}_{0}$ is the generalized diffusion coefficient of the physical
dimension $[D]=\mathrm{L}^{\sigma+2}/\mathrm{T}$, where $\mathrm{L}$ and $\mathrm{T}$
are units of the length and the time, respectively. In consequence, the non-linear diffusion
equation is as follows:
\begin{equation}
\frac{\partial}{\partial t}p(x,t)=D\frac{\partial}{\partial x}\left(p^{\sigma}(x,t)
\frac{\partial}{\partial x}p(x,t)\right).
\label{eq_3}
\end{equation}
A commonly known procedure for solving this class of equations is offered by the method of
similarity solutions that utilizes an algebraic symmetry of a differential equation. In order
to find its solution, we insert into Eq.~(\ref{eq_3}) a similarity transformation of the
algebraic form
\begin{equation}
p(x,t\!\mid\!x_{0})=\frac{1}{T(t)}F\left(\frac{x-x_{0}}{T(t)}\right)\equiv\frac{F(z)}{T(t)},
\;\;\text{with}\;\;z=\frac{x-x_{0}}{T(t)},
\label{eq_4}
\end{equation}
for the PDF of appearing a particle in $x$ at time $t$, if it was initially localized in the
position $x_{0}$ at time $t\!=\!0$. In this way, we effectively reduce the original partial
differential equation for the non-linear diffusion to the system of two ordinary differential
equations for the separate functions $T(t)$ and $F(z)$ which are relatively easy to solve.
We omit detailed calculations here and refer the interested reader to~\cite{Debn2012},
where the discussed method is accessibly explained. Thus, the final result takes the form
\begin{align}
p(x,t\!\mid\! x_{0})&=\frac{1}{T(t)}\left[a-\frac{b\sigma}{2D}\left(\frac{x-x_{0}}{T(t)}
\right)^{2}\right]^{\frac{1}{\sigma}},\;\;\text{with}\label{eq_5}\\
T(t)&=[b(\sigma+2)t]^{\frac{1}{\sigma+2}},\nonumber
\end{align}

\begin{figure}[t]
\centering
\includegraphics[scale=0.35]{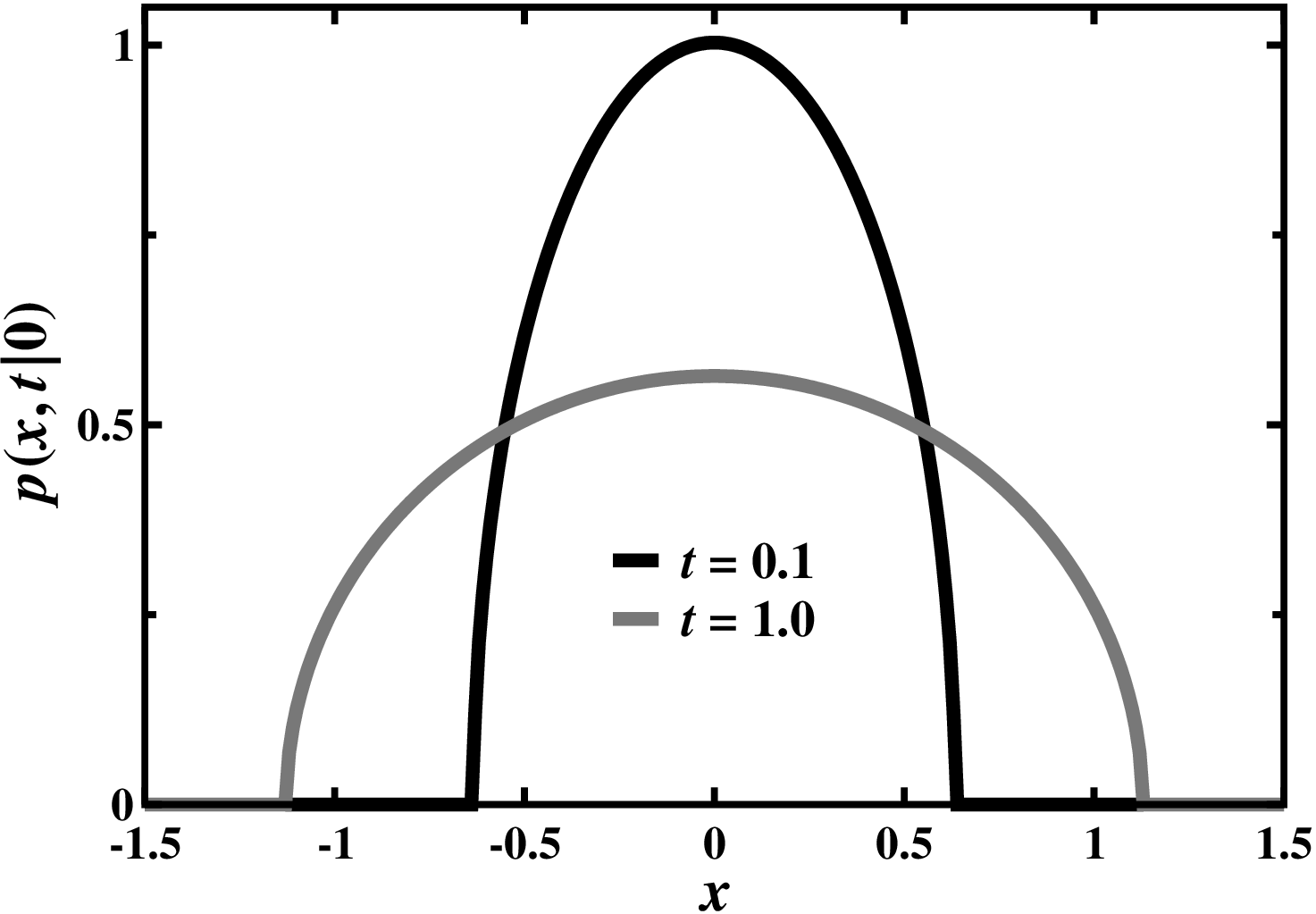}
\caption{Probability distribution function for the non-linear diffusion described by
Eq.~(\ref{eq_6}) in two consecutive moments of time. The value of the parameter $\sigma\!=\!2.0$
has been assumed and the diffusion coefficient $D\!=\!1.0$ has been established.}
\label{fig1}
\end{figure}
\noindent where $a$ and $b$ are arbitrary integration constants. Their specific values can be
determined by adopting suitable boundary conditions. For example, if we set $a\!=\!1$, with no
substantiation for now, and impose the normalization condition
$\int_{-\infty}^{\infty}p(x,t\!\mid\!x_{0})dx\!=\!1$ on Eq.~(\ref{eq_5}), performing
appropriate integration with a help of the Euler beta function
$\mathrm{B}(\nu,\mu)\!=\!\frac{\Gamma(\nu)\Gamma(\mu)}{\Gamma(\nu+\mu)}\!=\!
2\int_{0}^{1}z^{2\nu-1}(1-z^{2})^{\mu-1}dz$~\cite{Grad2007}, we find the unknown $b$ and
eventually a typical solution of Eq.~(\ref{eq_3}) in the Zel'dovitch-Barenblatt-Pattle
algebraic form~\cite{Zeld1958,Patt1959,Bare1972}
\begin{equation}
p(x,t\!\mid\!x_{0})=\frac{\mathcal{A}(\sigma)}{(D t)^{\frac{1}{\sigma+2}}}
\left[1-\mathcal{B}(\sigma)\frac{(x-x_{0})^{2}}{(D t)^{\frac{2}{\sigma+2}}}\right]
^{\frac{1}{\sigma}},
\label{eq_6}
\end{equation}
where the two $\sigma$-dependent coefficients in the above PDF are
\begin{equation}
\mathcal{A}(\sigma)=\left[\sqrt{\frac{\sigma}{2\pi (\sigma+2)}}\,
\frac{\Gamma\left(\frac{1}{\sigma}+\frac{3}{2}\right)}{
\Gamma\left(\frac{1}{\sigma}+1\right)}\right]^{\frac{2}{\sigma+2}}
\label{eq_7}
\end{equation}
and
\begin{equation}
\mathcal{B}(\sigma)=\sigma\left[\frac{1}{2(\sigma+2)}\right]^{\frac{2}{\sigma+2}}
\left[\frac{\sqrt{\pi}\,\Gamma\left(\frac{1}{\sigma}+1\right)}{\sqrt{\sigma}\,
\Gamma\left(\frac{1}{\sigma}+\frac{3}{2}\right)}\right]^{\frac{2\sigma}{\sigma+2}}.
\label{eq_8}
\end{equation}
The plot in Fig.~\ref{fig1} depicts profiles of the PDF in two different moments of time.
A supplementary comment is necessary at this point. The formulae given in Eqs.~(\ref{eq_5})
and (\ref{eq_6}) do not guarantee that the PDF for the non-linear diffusion is always
a real and non-negative function of $x$ as it should be by virtue of a very definition of the
probability density $p(x,t\!\mid\!x_{0})\geqslant0$. For this reason, we need to assume the
additional requirement that the PDF can only be determined on the finite support
$\lvert x-x_{0}\rvert\leqslant\mathcal{B}^{-\frac{1}{2}}(Dt)^{\frac{1}{\sigma+2}}$.
Everywhere outside this interval the PDF vanishes and such a property was taken into
account when performing integration in the normalization condition to figure out the
parameter $b$.

We are still aware choosing a value of the parameter $a\!=\!1$ without any explanation, which
may raise serious reservation. The argument behind such a choice is as follows. Taking the
limit $\sigma\!\to\!0$ and using the assertion that $\lim_{z\to\infty}\frac{\Gamma(z+\alpha)}
{\Gamma(z+\beta)}z^{\beta-\alpha}=1$ (see~\cite{Grad2007}) along with $\alpha=3/2$, $\beta=1$
and $z=1/\sigma$, we obtain from Eqs.~(\ref{eq_7}) and (\ref{eq_8}) that $\mathcal{A}\!=\!
1\!/\!\sqrt{4\pi}$ and $\mathcal{B}\!=\!\lim_{\sigma\to0}\sigma\!/\!4$. Simultaneously,
expressing the right-hand site of Eq.~(\ref{eq_6}) through the limit definition of the
exponential function $\mathrm{e}^{-z}=\lim_{n\to\infty}(1-\frac{z}{n})^{n}$ for $n\!=\!1/\sigma$,
we immediately retrieve the Gaussian distribution
\begin{equation}
p(x,t\!\mid\!x_{0})=\frac{1}{\sqrt{4\pi D t}}\exp\left[-\frac{(x-x_{0})^{2}}{4Dt}\right].
\label{eq_9}
\end{equation}
The above function is a fundamental solution of the linear partial differential equation
for the free diffusion given by Eq.~(\ref{eq_3}) with the initial condition $p(x,0\!\mid\!x_{0})
\!=\! \delta(x-x_{0})$, whenever $\sigma\!=\!0$. This result justifies our previous decision
to set $a\!=\!1$.

A solution of the partial differential equation is uniquely determined by imposing appropriate
boundary conditions. Among them the best known are periodic, reflecting and absorbing, as well
as a linear (weighted) combination of the last two boundary conditions. The latter are
a simplified version of the more general Robin boundary conditions, which assume that a given
function defined on the perimeter of a spacial domain, on which the solution of a partial
differential equation is to be found, corresponds to the weighted combination of this solution
and its first derivative over the spatial coordinate. Furthermore, the spacial boundary
conditions play a significant role in the context of the first-passage processes. A crucial
quantity related to this problem is the mean first-passage time (MFPT) or the mean hitting time
to a target. This average time is known to be finite when the process proceeds within the domain
confined by, for instance, reflecting boundary conditions, and diverges to infinity in an
unbounded space.  In order to calculate its value, we have to determine either the first-passage
time (FPT) distribution, the first moment of which is the MFPT, or the survival probability.
Both these functions are directly related to the PDF satisfying the absorbing boundary condition
at the target point. For this reason, the MFPT is sometimes called the mean time to absorption.
However, associating the absorbing boundary condition with the non-linear diffusion equation
rises a serious problem as we will show in the next section. Later, we will explain how to
overcome this obstacle in order to construct the basic quantities that quantitatively
characterise the first-passage properties of the non-linear diffusion. Finally, we present the
main results of this paper.

\section{\label{sec_3}Survival probability and first-passage time distribution}

Let us imagine a particle that starts from the initial position at $x\!=\!x_{0}\!>\!0$
and makes a diffusive motion along a semi-infinite interval $0\leqslant x<\infty$ with
a totally absorbing point at the origin $x\!=\!0$. What is a chance that the particle survives
before reaching the origin for the first time? The quantitative answer to this question
is given in terms of a survival probability $Q(t\!\mid\!x_{0})$. In general, it is defined
as a spatial integral of the PDF over a certain region of space where a stochastic process
takes place in the presence of an absorbing trap localized somewhere at the perimeter or inside
of this region~\cite{Gard2004}. For diffusion in the semi-infinite interval we have
\begin{equation}
Q(t\!\mid\!x_{0})=\int_{0}^{\infty}p(x,t\!\mid\!x_{0})\,\mathrm{d}x,
\label{eq_10}
\end{equation}
where the PDF is a solution of a partial differential equation which satisfies the
absorbing boundary condition, i.e. $p(0,t\!\mid\!x_{0})\!=\!0$ at the origin $x\!=\!0$.
Furthermore, the survival probability is supplemented by additional conditions resulting
from natural requirements imposed on the PDF. The first property is a direct consequence of
an initial condition $p(x,0\!\mid\!x_{0})\!=\!\delta(x-x_{0})$ stating that the particle begins
its diffusive motion from the position localized at $x\!=\!x_{0}$. Using this property in
Eq.~(\ref{eq_10}) along with the normalization condition of the Dirac delta function, i.e.
$\int_{0}^{\infty}\delta(x-x_{0})\,\mathrm{d}x\!=\!1$, we easily obtain that
$Q(0\!\mid\!x_{0})\!=\!1$. The second feature of the survival probability relates to the
situation when the initial position of a particle coincides with the absorbing point, i.e.
$x_{0}\!=\!0$. In this case the particle remains there forever which means that
$Q(t\!\mid\!0)\!=\!0$ at any time $t\!>\!0$. This rule also holds for the PDF. The two
properties of the survival probability considered so far are mostly used together with the so
called backward diffusion (Fokker-Planck) equation describing the time evolution of this
quantity~\cite{Risk1989}. The last property of the survival probability emerges form our
conviction that the particle will be eventually absorbed at the origin for times large enough, so
$Q(t\!\mid\!x_{0})\!\to\!0$ when $t\!\to\!\infty$. We should, however, emphasize that
this asymptotic limit is not always satisfied. A good example is a biased diffusion in the
semi-infinite interval where the behavior of the survival probability depends on whether
a drift velocity is positive or negative (see~\cite{Redn2001}).

Armed with the survival probability, we can now consider of how long the diffusing particle
will persevere in the semi-infinite interval before reaching the absorbing target for the first
time. For this purpose, one needs to calculate the first derivative of the cumulative probability
function $1\!-\!Q(t\!\mid\!x_{0})$ with respect to time, which gives
\begin{equation}
\frac{\mathrm{d}}{\mathrm{d}t}Q(t\!\mid\!x_{0})=-F(t\!\mid\!x_{0}).
\label{eq_11}
\end{equation}
The function $F(t\!\mid\!x_{0})$ specifies the FPT distribution and its first moment determines
the MFPT from the initial position at $x\!=\!x_{0}$ to the target localized at the origin
$x\!=\!0$:
\begin{equation}
\mathcal{T}(x_{0})=\int_{0}^{\infty}t\,F(t\!\mid\!x_{0})\,\mathrm{d}t.
\label{eq_12}
\end{equation}
Alternatively, inserting Eq.~(\ref{eq_11}) into the above formula and performing an integration
{\it per partes} under the conditions $Q(0\!\mid\!x_{0})\!=\!1$ and $Q(\infty\!\mid\!x_{0})
\!=\!0$, we obtain
\begin{equation}
\mathcal{T}(x_{0})=-\int_{0}^{\infty}t\,\mathrm{d}Q(t\!\mid\!x_{0})=\int_{0}^{\infty}
Q(t\!\mid\!x_{0})\,\mathrm{d}t.
\label{eq_13}
\end{equation}
There are two additional properties regarding the FPT distribution. The first property
corresponds to the statement that this density function is by definition normalized to unity.
To show this, we have to begin with the integration of Eq.~(\ref{eq_11}) over the time
variable in the range from $0$ to $t$. The result is as follows:
\begin{equation}
Q(t\!\mid\!x_{0})=1-\int_{0}^{t}F(\tau\!\mid\!x_{0})\,\mathrm{d}\tau,
\label{eq_14}
\end{equation}
where the property $Q(0\!\mid\!x_{0})\!=\!1$ has been exploited. We will utilize this important
equation in Sec.~\ref{sec_4_1}. On the other hand, demanding that $t\!\to\infty$ and knowing
that $Q(\infty\!\mid\!x_{0})\!=\!0$, we readily obtain from Eq.~(\ref{eq_11}) the required
normalization condition
\begin{equation}
\int_{0}^{\infty}F(t\!\mid\!x_{0})\,\mathrm{d}t=1.
\label{eq_15}
\end{equation}
This condition implicates the particle is sure to hit the absorbing point, although the mean
time, by which such an event occurs, does not necessarily be finite. The second property
refers to the relationship between the FPT distribution and the probability current (flux)
$j(x,t\!\mid\!x_{0})$. The latter quantity is, in turn, related to the PDF through the conserved
current relation, which is expressed by the continuity equation
\begin{equation}
\frac{\partial}{\partial t}p(t,x\!\mid\!x_{0})+\frac{\partial}{\partial x}
j(x,t\!\mid\!x_{0})=0.
\label{eq_16}
\end{equation}
A combination of this equation with the first derivative of Eq.~(\ref{eq_10}) with respect to
time gives that
\begin{equation}
\frac{\mathrm{d}}{\mathrm{d}t}Q(t\!\mid\!x_{0})=\int_{0}^{\infty}\frac{\partial}
{\partial t}p(x,t\!\mid\!x_{0})\,\mathrm{d}x=-\int_{0}^{\infty}\mathrm{d}j(x,t\!\mid\!x_{0})=
j(0,t\!\mid\!x_{0}),
\label{eq_17}
\end{equation}
where we have assumed that the current $j(x,t\!\mid\!x_{0})\!\to\!0$ for $x\!\to\!\infty$ and
the non-zero current $j(0,t\!\mid\!x_{0})$ determines the rate of absorption at the point
$x\!=\!0$. On the basis of Eqs.~(\ref{eq_11}) and (\ref{eq_17}) we obtain the second property
for the FPT distribution
\begin{equation}
F(t\!\mid\!x_{0})=-j(0,t\!\mid\!x_{0}).
\label{eq_18}
\end{equation}
This formula allows us to derive the FPT distribution for diffusion in the semi-infinite
interval directly from the probability current at the origin. But firstly, we have to
solve the corresponding partial differential equation for the PDF with the absorbing boundary
condition also imposed at the origin. In what follows, we demonstrate that such a procedure is
possible for the linear diffusion equation, while not feasible in the case of the non-linear
diffusion equation.

\subsection{\label{sec_31}The linear diffusion equation}

As we have shown in Sec.~\ref{sec_2}, a typical solution of the linear partial differential
equation for the free diffusion is the Gaussian PDF given by Eq.~(\ref{eq_9}). A conventional
technique for solving this type of differential equation in the presence of the absorbing point
is the image method. Let us clarify that this familiar method emerges form a more general theory
of Green's functions and found successful application also in electrostatics. The idea consists
in a creation of a virtual system making up of the particle initially in the position
$x\!=\!x_{0}\!>\!0$ and the fictitious "antiparticle" located in the position
$x\!=\!-x_{0}\!<\!0$. When the particle begins to diffuse in the semi-infinite interval
$[0,\infty)$, then the antiparticle does the same like the mirror image in the semi-infinite
interval $(-\infty,0]$. The free diffusion proceeds until both the particles meet for the first
time at the origin, where they disappear due to "annihilation". Owing to the linearity of the
diffusion equation, the resulting PDF
\begin{equation}
p(x,t\!\mid\!x_{0})=\frac{1}{\sqrt{4\pi D t}}\left\{\exp\left[-\frac{(x-x_{0})^{2}}{4Dt}\right]-
\exp\left[-\frac{(x+x_{0})^{2}}{4Dt}\right]\right\}
\label{eq_19}
\end{equation}
is the combination of two Gaussian distributions and satisfies the absorbing boundary condition
$p(0,t\!\mid\!x_{0})\!=\!0$. On the other hand, a construction of the ordinary diffusion equation
(see Eq.~(\ref{eq_3}) with $\sigma\!=\!0$) on the basis of Eq.~(\ref{eq_16}) requires that the
probability current must be of the following form:
\begin{equation}
j(x,t\!\mid\!x_{0})=-D\frac{\partial}{\partial x}p(x,t\!\mid\!x_{0}).
\label{eq_20}
\end{equation}
Hence, calculating the first derivative of the PDF in Eq.~(\ref{eq_19}) with respect to $x$
and setting $x\!=\!0$, we show that the FPT distribution given by Eq.~(\ref{eq_18}) is as
follows:
\begin{equation}
F(t\!\mid\!x_{0})=\frac{x_{0}}{\sqrt{4\pi D\,t^{3}}}\exp\left(-\frac{x_{0}^{2}}
{4Dt}\right).
\label{eq_21}
\end{equation}
In the long-time limit $\sqrt{Dt}\!\gg\!x_{0}$, for which the diffusion length is much grater
than the initial distance to the origin, the above function reduces to
$F(t\!\mid\!x_{0})\!\sim\!x_{0}\,t^{-3/2}$. The existence of this long time tail makes the
MFPT from $x_{0}$ to the origin infinite, because according to Eq.~(\ref{eq_12})
$\mathcal{T}(x_{0})\!\sim\!\int_{0}^{\infty}t\times t^{-3/2}\,\mathrm{d}t\!=\!\infty$. On the
other hand, the FPT distribution in Eq.~(\ref{eq_21}) fulfills the normalization condition
embodied by Eq.~(\ref{eq_15}) which means that the diffusing particle is paradoxically sure to
return to the origin. As a consequence of this statement, we can conclude that the survival
probability of the particle should vanish in the long time limit. Indeed, inserting
Eq.~(\ref{eq_21}) into Eq.~(\ref{eq_14}) and utilizing the integral
$\int_{0}^{t}\tau^{-3/2}\exp\left(-\frac{\alpha}{\tau}\right)\mathrm{d}\tau\!=\!\sqrt{\frac{\pi}
{\alpha}}\,\mathrm{erfc}\left(\sqrt{\frac{\alpha}{t}}\right)$~\cite{Grad2007}, we finally obtain
that
\begin{equation}
Q(t\!\mid\!x_{0})=1-\mathrm{erfc}\left(\frac{x_{0}}{\sqrt{4Dt}}\right)=
\mathrm{erf}\left(\frac{x_{0}}{\sqrt{4Dt}}\right),
\label{eq_22}
\end{equation}
where $\mathrm{erfc}(z)\!=\!1-\mathrm{erf}(z)$ is the complementary error function, whereas
$\mathrm{erf}(z)$ stands for the error function. This result indicates that the survival
probability for the ordinary diffusion in the semi-infinite interval with the absorbing barrier
at the origin tends to zero in the long-time limit, i.e. $Q(t\!\mid\!x_{0})\!\to\!0$ if
$t\!\to\!\infty$. This property arises from the Taylor expansion of the error function
$\mathrm{erf}(z)\!\sim\!\frac{2z\exp\left(-z^{2}\right)}{\sqrt{\pi}}$ for $z\!\to\!0$.
The second property of the survival probability, namely that $Q(0\!\mid\!x_{0})\!=\!1$, is also
satisfied. To show this, it is enough to use the asymptotic representation of the error function
$\mathrm{erf}(z)\!\sim\!1-\frac{\exp\left(-z^{2}\right)}{\sqrt{\pi}z}$ for $z\!\to\!\infty$.

\subsection{\label{sec_32}The non-linear diffusion equation}

The image method allows us to represent the solution of the ordinary diffusion equation in the
presence of the absorbing trap as the superposition of probability and "anti"-probability
density functions. However, the linear combination of the PDFs can not be the solution of the
non-linear diffusion equation displayed in Eq.~(\ref{eq_3}), so we cannot use the image method
in this case. For this reason, we will proceed as Wang et al, who analysed the first-passage
problem for a more general non-linear diffusion equation~\cite{Wang2008}. Equation~(\ref{eq_3})
is a simplified version of that fractional and heterogeneous differential equation.

To begin with, let us consider the non-linear diffusion in the semi-infinite interval and
establish the absorbing boundary condition $p(0,t\!\mid\!x_{0})\!=\!0$ at the origin. Then
applying this condition directly to the PDF in Eq.~(\ref{eq_5}), we readily find that
\begin{equation}
p(x,t\!\mid\!x_{0})=\left[\frac{\sigma x(2x_{0}-x)}{2D(\sigma+2)t}\right]^{\frac{1}{\sigma}}.
\label{eq_23}
\end{equation}
Having this result to our disposal, we can now determine the appropriate expression for the
survival probability given by Eq.~(\ref{eq_10}). An astute look at Eq.~(\ref{eq_23})
convinces us that this PDF can however take negative values and even be imaginary for some
values of the parameter $\sigma$. On the other site, the integral defining the survival
probability becomes divergent when substituting the PDF of the algebraic form. To be sure the
PDF is real and non-negative quantity and the integral over a space variable $x$ is convergent,
one needs to multiply the PDF by the Heviside unit step function $\Theta(2x_{0}-x)$ equal to 1,
if $x\leqslant2x_{0}$, and 0, if $x>2x_{0}$. The result of this operation is as follows:
\begin{equation}
Q(t\!\mid\!x_{0})=\int_{0}^{\infty}\Theta(2x_{0}-x)\,p(x,t\!\mid\!x_{0})\,\mathrm{d}x=
\int_{0}^{2x_{0}}p(x,t\!\mid\!x_{0})\,\mathrm{d}x.
\label{eq_24}
\end{equation}
In the last step, it is enough to put the PDF in Eq.~(\ref{eq_23}) into the above formula and
use the integral representation of the Euler beta function, $\mathrm{B}(\nu,\mu)=\int_{0}^{1}
z^{\nu-1}(1-z)^{\mu-1}\mathrm{d}z$ \cite{Grad2007}. In this way, we show that the survival probability
\begin{equation}
Q(t\!\mid\!x_{0})=2
\mathrm{B}\left(\frac{1}{\sigma},\frac{1}{2}\right)
\left[\frac{\sigma x_{0}^{\sigma+2}}{2D(\sigma+2)^{\sigma+1}t}\right]^{\frac{1}{\sigma}}.
\label{eq_25}
\end{equation}
According to Eq.~(\ref{eq_11}) the first derivative of the survival probability in
Eq.~(\ref{eq_25}) with respect to time results in the FPT distribution
\begin{equation}
F(t\!\mid\!x_{0})=2
\mathrm{B}\left(\frac{1}{\sigma},\frac{1}{2}\right)
\left[\frac{\sigma^{2}x_{0}^{\sigma+2}}{2D(\sigma(\sigma+2)t)^{\sigma+1}}\right]^{\frac{1}
{\sigma}}.
\label{eq_26}
\end{equation}
The survival probability given by Eq.~(\ref{eq_25}) and the FPT distribution given by
Eq.~(\ref{eq_26}), the both of the algebraic form, when inserted, respectively, in
Eq.~(\ref{eq_13}) and Eq.~(\ref{eq_12}), make the MFPT to the origin to be divergent for the
non-linear diffusion. The same result obviously holds for the ordinary diffusion.
 
Nevertheless, let us briefly consider whether all the formulas derived so far are really correct.
The functions embodied by Eqs.~(\ref{eq_25}) and (\ref{eq_26}) undoubtedly tend to zero in the
long-time limit and are zero for any time when the initial position of the particle coincides
with the origin, i.e. $Q(t\!\mid\!0)\!=\!0$ and $F(t\!\mid\!0)\!=\!0$. But, it is clearly
evident that the survival probability does not satisfy the initial condition
$Q(0\!\mid\!x_{0})\!=\!1$. Wang et al did not verify this crucial property in
Ref.~\cite{Wang2008}. In turn, the FPT distribution in Eq.~(\ref{eq_26}) cannot be normalized
to unity because the integral expression in Eq.~(\ref{eq_15}) is divergent for any value of the
parameter $\sigma$. Therefore, what is the main reason that these two fundamental properties of
the survival probability and the FPT distribution are broken?

The key to unravel this riddle is the absorbing boundary condition imposed on the non-linear
diffusion equation. To show this, we need to compare Eq.~(\ref{eq_3}) with the conserved current
relation given by Eq.~(\ref{eq_16}). Hence, the probability current is
\begin{equation}
j(x,t\!\mid\!x_{0})=-D\,p^{\sigma}(x,t\!\mid\!x_{0})\frac{\partial}{\partial x}
p(x,t\!\mid\!x_{0}).
\label{eq_27}
\end{equation}
Taking advantage of the above equation along with Eq.~(\ref{eq_18}) enables immediate
determination of the FPT distribution. In the physical sense, the probability current defines
an appropriate measure of the absorption rate. We can easily check, utilizing Eqs.~(\ref{eq_19})
and (\ref{eq_20}), that the rate of absorption at the point $x\!=\!0$ is non-zero for the
ordinary diffusion. However, the case of the non-linear diffusion equation does not reveal such
a behavior. Indeed, by inserting Eq.~(\ref{eq_23}) in Eq.~(\ref{eq_27}), we obtain that
\begin{equation}
j(x,t\!\mid\!x_{0})=\frac{\lvert x-x_{0}\rvert}{(\sigma+2)t}\left[\frac{\sigma x(2x_{0}-x)}
{2D(\sigma+2)t}\right]^{\frac{1}{\sigma}}=\frac{\lvert x-x_{0}\rvert}{(\sigma+2)t}\,
p(x,t\!\mid\!x_{0}),
\label{eq_28}
\end{equation}
and because $p(0,t\!\mid\!x_{0})\!=\!0$ at the absorbing point, so does the probability
current $j(0,t\!\mid\!x_{0})\!=\!0$. But, the disappearance of this current (flux) means
the presence of the reflecting and not absorbing boundary condition at that point. Thus,
does this point absorb or reflect the diffusing particle? We are not able to dispel this
ambiguity unequivocally. Nevertheless, our analysis shows that the method applied by Wang et
al in Ref.~\cite{Wang2008} is inappropriate and should not be used to explore the first-passage
properties of the non-linear diffusion equation.

\section{\label{sec_4}First-passage properties of non-linear diffusion}

\subsection{\label{sec_4_1}General method}

Fortunately, the situation outlined in the previous section is not completely hopeless.
In essence, there exists an alternative framework thanks to which the solution of the
first-passage problem can successfully be achieved even in the case of the non-linear
diffusion. To show this, we will continue our study of the non-linear diffusion in the
semi-infinite interval $[0,\infty)$ in the proceeding sections.

The method that is at our disposal does not suppose the existence of the absorbing barrier
at the target point $x\!=\!0$. Instead, it treats this point as a "safe marina" to which
the particle returns many times after the first visit. To be more precise, the method
utilizes a duo of coupled equations. The first equation, we have already met in
Eq.~(\ref{eq_14}), constitutes a relation between the survival probability
$Q(t\!\mid\!x_{0})$ and the FPT distribution $F(t\!\mid\!x_{0})$. The second relation
combines the FPT distribution with the PDFs and has the form of the integral equation
\begin{equation}
p(0,t\!\mid\!x_{0})=\int_{0}^{t}F(\tau\!\mid\!x_{0})\,p(0,t-\tau\!\mid\!0)\,\mathrm{d}\tau.
\label{eq_29}
\end{equation}
This equation defines the PDF or more precisely the propagator from $x_{0}$ to the target at
$x\!=\!0$ for any stochastic dynamics as an integral over the first time to reach the point $0$
at a time $\tau\!\leqslant\!t$ followed by a loop from $(0,\tau)$ to $(0,t)$ in the remaining
time $t\!-\!\tau$~\cite{Krap2010}. Note the integral expression in Eq.~(\ref{eq_29}) is a time
convolution of two distribution functions, thus a price we must pay to determine the FPT
distribution $F(t\!\mid\!x_{0})$ involves the use of the Laplace transformation. The convolution
theorem states that the Laplace transformation, defined as
$\tilde{f}(s)\!=\!\mathcal{L}[f(t);t]\!:=\!\int_{0}^{\infty}f(t)\mathrm{e}^{-st}\mathrm{d}t$,
of the convolution $f(t)\!\ast\!g(t)\!:=\!\int_{0}^{t}f(\tau)g(t-\tau)\mathrm{d}\tau$ of two
integrable functions $f(t)$ and $g(t)$ is the product of their Laplace transforms, i.e.
$\mathcal{L}[f(t)\!\ast\!g(t);t]\!=\!\tilde{f}(s)\tilde{g}(s)$~\cite{Schi1999}. Therefore, we
can convert Eq.~(\ref{eq_29}) into the algebraic form
\begin{equation}
\tilde{F}(s\!\mid\!x_{0})=\frac{\tilde{p}(0,s\!\mid\!x_{0})}{\tilde{p}(0,s\!\mid\!0)}.
\label{eq_30}
\end{equation}
In turn, performing the Laplace transformation of Eq.~(\ref{eq_14}) yields
\begin{equation}
\tilde{Q}(s\!\mid\!x_{0})=\frac{1}{s}\left[1-\tilde{F}(s\!\mid\!x_{0})\right].
\label{eq_31}
\end{equation}
The combination of the last two expressions makes the direct relationship between the
survival probability and the PDFs in the Laplace domain:
\begin{equation}
\tilde{Q}(s\!\mid\!x_{0})=\frac{1}{s}\left[1-\frac{\tilde{p}(0,s\!\mid\!x_{0})}
{\tilde{p}(0,s\!\mid\!0)}\right],
\label{eq_32}
\end{equation}
Armed with the above equation and Eq.~(\ref{eq_13}), we can calculate the MFPT
\begin{equation}
\mathcal{T}(x_{0})=\lim_{s\to0}\int_{0}^{\infty}Q(t\!\mid\!x_{0})\mathrm{e}^{-st}\,\mathrm{d}t
=\lim_{s\to0}\tilde{Q}(s\!\mid\!x_{0})
\label{eq_33}
\end{equation}
from the initial position $x_{0}$ to the origin of the semi-infinite interval. On the other
hand, carrying out the inverse Laplace transformation of Eq.~(\ref{eq_32}), which usually is
not trivial operation, allows one to find the survival probability and hence the FPT
distribution (see Eq.~(\ref{eq_11})) in the real space. This is shown in the next section
where we will obtain the exact results for the non-linear diffusion with the exceptional
values of the parameter $\sigma\!=\!1$ and $2$, while the approximate formula will be derived
for any values of $\sigma$ falling in-between.

\subsection{\label{sec_4_2}Results for non-linear diffusion}

The key quantity appearing in Eqs.~\ref{eq_30} and \ref{eq_32} is the Laplace transform
of the PDF, i.e. the propagator $p(0,t\!\mid\!x')$ for the free non-linear diffusion from
$x'$ to the origin at $x\!=\!0$, where $x'$ corresponds to the initial $x_{0}$ or the final
position (the target) $0$. In the latter case the propagator stands for the so-called return
probability density. From the very beginning we posit that the parameter $\sigma$ in
Eq.~(\ref{eq_2}) is assumed to be non-negative. By virtue of this condition, the two
$\sigma$-dependent coefficients in Eq.~(\ref{eq_6}), namely $\mathcal{A}(\sigma)$ and
$\mathcal{B}(\sigma)$ (see Eqs.~(\ref{eq_7}) and (\ref{eq_8}), respectively), are also
non-negative and real. In addition, the following inequality, namely
$Dt\geqslant(\sqrt{\mathcal{B}(\sigma)}\mid\!x-x_{0}\!\mid)^{\sigma+2}$, guarantees
that the PDF in Eq.~(\ref{eq_6}) is non-negative and real as the function of time. This
necessary condition can be formally expressed through the Heviside unit step function
$\Theta(z)$ that for $z\!\geqslant\!0$ equals to 1 and 0 if $z\!<\!0$. Because the same
requirement must concern the propagator, we conclude that
\begin{equation}
p(0,t\!\mid\!x')=\Theta\!\left(t-D^{-1}(\sqrt{\mathcal{B}(\sigma)}\,\lvert x'\rvert)^
{\sigma+2}\right)\frac{\mathcal{A}(\sigma)}{(Dt)^{\frac{1}{\sigma+2}}}
\left[1-\mathcal{B}(\sigma)\frac{x'^{\,2}}{(D t)^{\frac{2}{\sigma+2}}}\right]
^{\frac{1}{\sigma}}.
\label{eq_34}
\end{equation}
Consequently, the Laplace transform of the propagator, which is defined as
$\tilde{p}(0,s\!\mid\!x')\!=\!\int_{0}^{\infty}p(0,t\!\mid\!x')\mathrm{e}^{-st}
\mathrm{d}t$, takes the following form:
\begin{equation}
\tilde{p}(0,s\!\mid\!x')=\int_{\eta(x',\sigma)}^{\infty}\frac{\mathcal{A}(\sigma)}
{(D t)^{\frac{1}{\sigma+2}}}\left[1-\left(\frac{\eta(x',\sigma)}{t}\right)^{\frac{2}
{\sigma+2}}\right]^{\frac{1}{\sigma}}\exp(-st)\,\mathrm{d}t,
\label{eq_35}
\end{equation}
where the auxiliary function $\eta(x',\sigma)\!=\!D^{-1}(\sqrt{\mathcal{B}(\sigma)}\,\lvert x'
\rvert)^{\sigma+2}$ allows us to use the shorthand notation in the integrand and the lower
limit of the integral. The formula displayed in Eq.~(\ref{eq_35}) constitutes the starting
point for studies of the first-passage time properties of the non-linear diffusion. However,
a precise calculation of the integral appearing in Eq.~(\ref{eq_35}) poses a great challenge
whenever arbitrary values of the parameter $\sigma\!>\!0$ are taken into account. Nevertheless,
there exist, at least, three exceptions when this operation can be precisely performed.
To continue, we will consider them first.

\subsubsection{\label{sec_4_2_1}Exact results}

The first exception that corresponds to the parameter $\sigma\!=\!\frac{1}{N}$, where $N$
represents any natural number, is rather trivial and we skip its analysis in the present
paper. The next case is relatively simple and refers to the parameter $\sigma\!=\!1$.
Here, the coefficient $\mathcal{A}(1)\!=\!\left(\frac{3}{32}\right)^{1/3}$ and the lower
limit of integration $\eta(x',1)\!=\!\frac{2\lvert x'\rvert^{3}}{9D}$. Therefore, the Laplace
transform of the propagator in Eq.~(\ref{eq_35}) reads
\begin{equation}
\tilde{p}(0,s\!\mid\!x')=\left(\frac{3}{32}\right)^{\frac{1}{3}}(\sqrt{D}s)^{-\frac{2}{3}}
\left[\Gamma\left(\frac{2}{3},\frac{2\lvert x'\rvert^{3}}{9D}s\right)+
\left(\frac{2s}{9D}\right)^{\frac{2}{3}}x'^{\,2}\,\mathrm{Ei}\left(-\frac{2\lvert x'\rvert^{3}}
{9D}s\right)\right],
\label{eq_36}
\end{equation}
where $\Gamma(\alpha,z)\!=\!\int_{z}^{\infty}u^{\alpha-1}\mathrm{e}^{-u}\mathrm{d}u$ is the upper
incomplete gamma function and $\mathrm{Ei}(-z)\!=\!-\int_{z}^{\infty}u^{-1}\mathrm{e}^{-u}
\mathrm{d}u$ stands for the exponential integral function~\cite{Grad2007}. Let us note that when
$x'\!=\!0$ the above propagator operating in the Laplace domain becomes the return probability
density $\tilde{p}(0,s\!\mid\!0)=\left(\frac{3}{32}\right)^{\frac{1}{3}}\Gamma\left(\frac{2}{3}
\right)(\sqrt{D}s)^{-\frac{2}{3}}$. This formula emerges from the fact that $\Gamma(z,0)\!=\!
\Gamma(z)$ and the assertion stating that $\lim_{z\to0}z^{2}\mathrm{Ei}(-az^{3})\!=\!0$. The
latter property is easy verified by making use of the L'Hospital theorem. The aforementioned
return probability density can also be obtained from a direct integration in Eq.~(\ref{eq_35})
with $\eta(0,1)\!=\!0$.

In contrast to the previous two exceptions, a slightly more difficult is the case when the
parameter $\sigma\!=\!2$. Here, we only exhibit the final expression for the Laplace
transform of the non-linear diffusion propagator with such a parameter. Using the procedure
sketched in Appendix~A, we readily infer from Eq.~\ref{eq_35} that
\begin{equation}
\tilde{p}(0,s\!\mid\!x')=\frac{\pi\lvert x'\rvert^{3}}{8\sqrt{2}D}\exp\!\left(-
\frac{\pi^{2}x'^{\,4}}{32D}s\right)\left[\mathrm{K}_{\frac{3}{4}}\!\left(\frac{
\pi^{2}x'^{\,4}}{32D}s\right)-\mathrm{K}_{\frac{1}{4}}\!\left(\frac{\pi^{2}x'^{\,4}}{32D}s
\right)\right],
\label{eq_37}
\end{equation}
\newpage where $\mathrm{K}_{\nu}(z)$ is the modified Bessel function of the second kind~\cite{Grad2007}.
We also prove there that the Laplace transform of the return probability density $\tilde{p}(0,s\!\mid\!0)\!=\!\Gamma\left(\frac{3}{4}\right)(\pi\sqrt{D})^{-1/2}\,s^{-3/4}$.
\begin{figure}[t]
\centering
\includegraphics[scale=0.35]{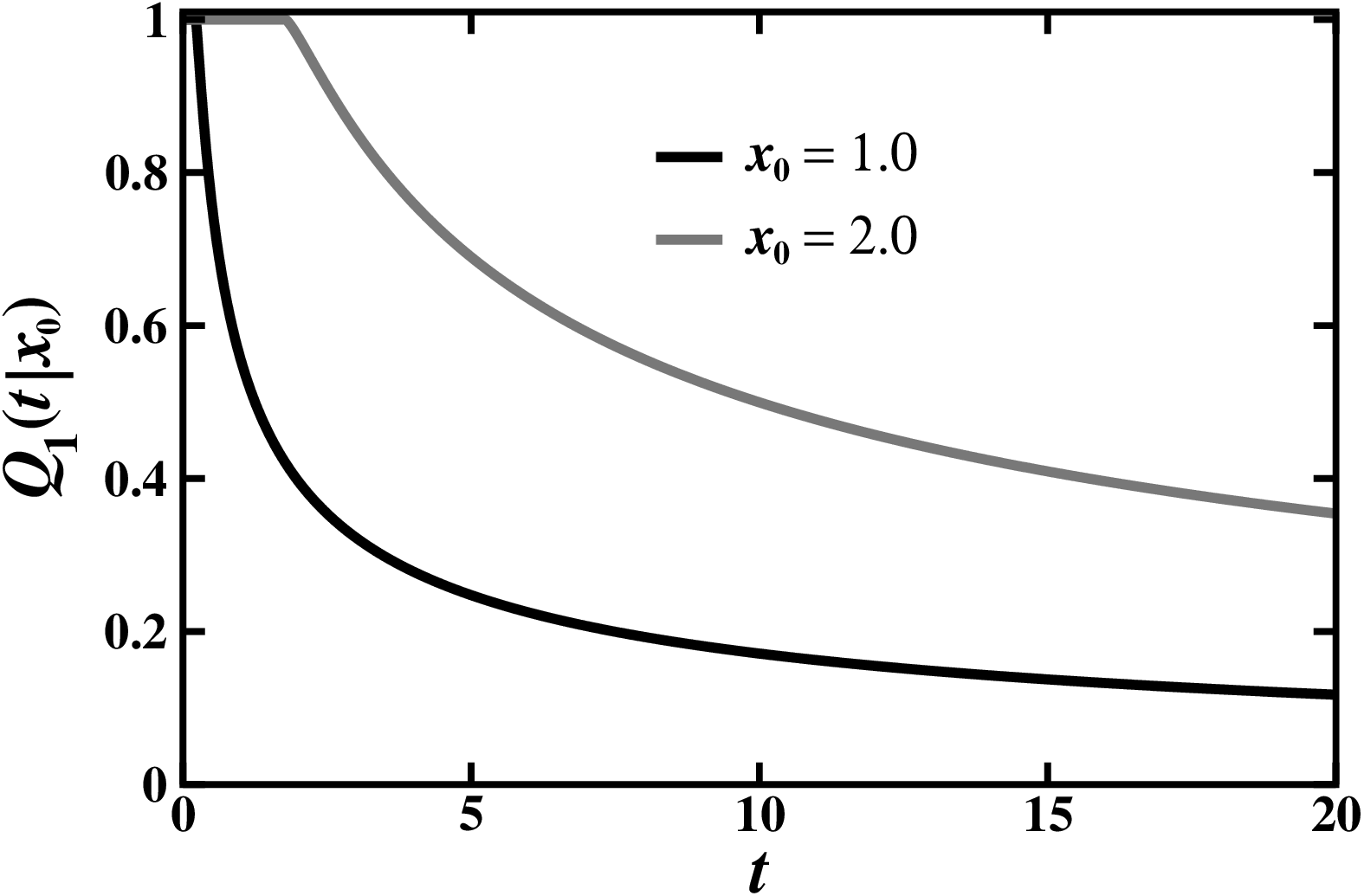}
\caption{Time course of the survival probability given by Eq.~(\ref{eq_40}) for the non-linear
diffusion with the parameter $\sigma\!=\!1$. Here, two values of the initial position $x_{0}$
relative to the target point located at the origin $x\!=\!0$ of the semi-infinite interval have
been chosen and the value of the diffusion coefficient $D\!=\!1.0$ has been assumed.}
\label{fig2}
\end{figure}
\begin{figure}[b]
\centering
\includegraphics[scale=0.3]{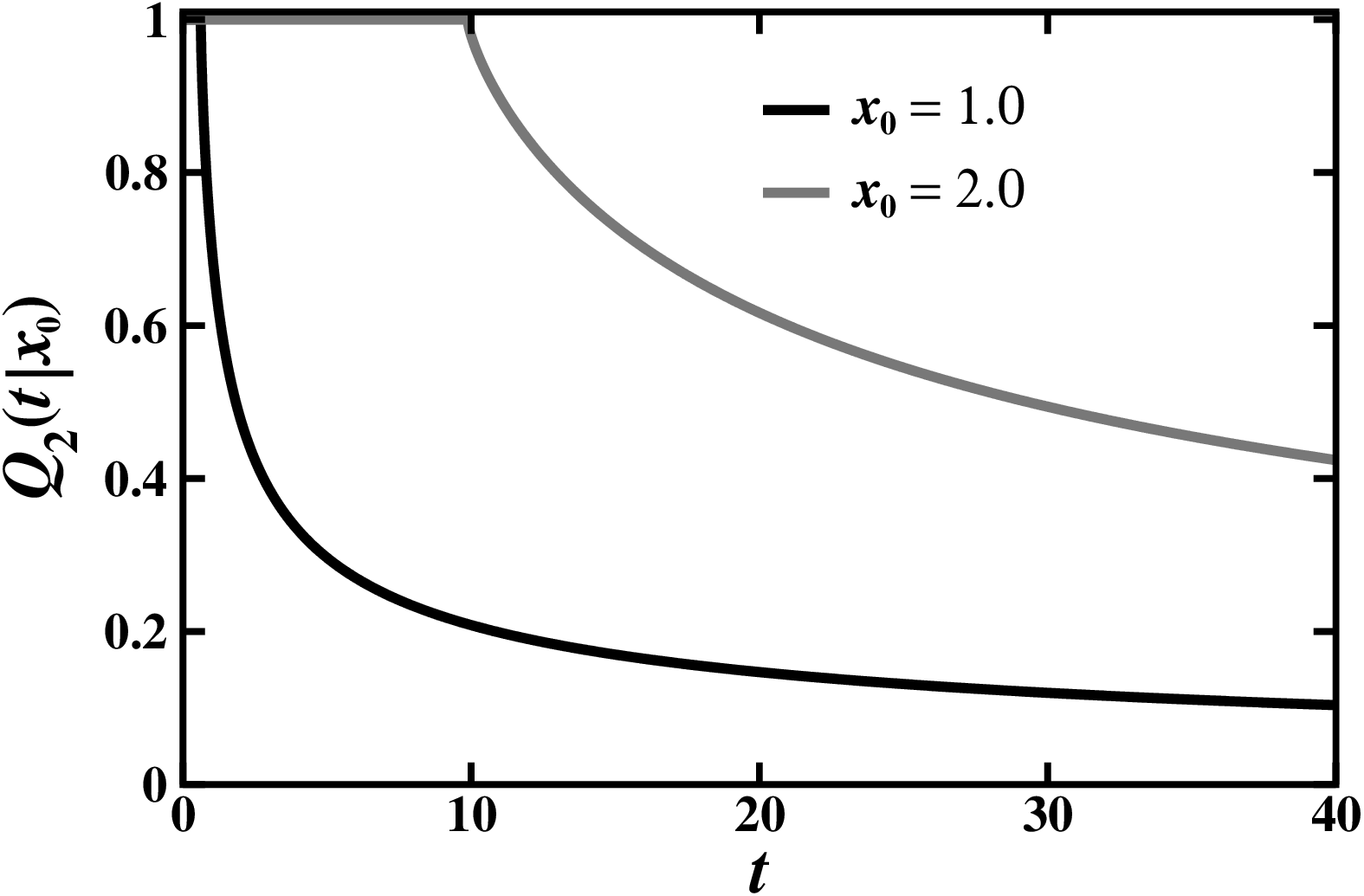}
\caption{Time course of the survival probability given by Eq.~(\ref{eq_41}) for the non-linear
diffusion with the parameter $\sigma\!=\!2$. Here, two values of the initial position $x_{0}$
relative to the target point located at the origin $x\!=\!0$ of the semi-infinite interval have
been chosen and the value of the diffusion coefficient $D\!=\!1.0$ has been assumed.}
\label{fig3}
\end{figure}

Given the Laplace transforms of the propagators in Eqs.~(\ref{eq_36}) and (\ref{eq_37}) with
$x'\!=\!x_{0}$, as well as the corresponding return probability densities with $x'\!=\!0$,
we are now set to determine the Laplace transforms of the survival probabilities for
the non-linear diffusion indexed by the power-law exponents $\sigma\!=\!1$ and $2$ (see
Eq.~(\ref{eq_3})). For this purpose, it is enough to insert each of these formulae into
Eq.~(\ref{eq_32}) and straightforwardly obtain the Laplace transform of the survival probability
for $\sigma\!=\!1$:
\begin{equation}
\tilde{Q}_{1}(s\!\mid\!x_{0})=\frac{1}{s}-\frac{1}{\Gamma\left(\frac{2}{3}\right)}
\left[\frac{1}{s}\,\Gamma\left(\frac{2}{3},\frac{2\lvert x_{0}\rvert^{3}}{9D}s\right)-
\left(\frac{2\lvert x_{0}\rvert^{3}}{9D}\right)^{\frac{2}{3}}\frac{1}{s^{2/3}}\,
\Gamma\left(0,\frac{2\lvert x_{0}\rvert^{3}}{9D}s\right)\right],
\label{eq_38}
\end{equation}
where the exponential integral function has been replaced with the upper incomplete gamma
function in accordance with the relation $\mathrm{Ei}(-z)\!=\!-\Gamma(0,z)$~\cite{Grad2007},
and the Laplace transform of the survival probability for $\sigma\!=\!2$:
\begin{equation}
\tilde{Q}_{2}(s\!\mid\!x_{0})=\frac{1}{s}-\frac{\pi^{3/2}\lvert x_{0}\rvert^{3}}{8\sqrt{2}\,
D^{3/4}\,\Gamma\left(\frac{3}{4}\right)s^{1/4}}\exp\!\left(-\frac{\pi^{2}x_{0}^{\,4}}{32D}s
\right)\left[\mathrm{K}_{\frac{3}{4}}\!\left(\frac{\pi^{2}x_{0}^{\,4}}{32D}s\right)
-\mathrm{K}_{\frac{1}{4}}\!\left(\frac{\pi^{2}x_{0}^{\,4}}{32D}s\right)\right].
\label{eq_39}
\end{equation}
At first glance, the above two expressions appear to be quite intricate regarding the
performance of the inverse Laplace transform. However, taking advantage of Eqs.~(\ref{eq_B9})
and (\ref{eq_B10}) derived in Appendix~B, we convince ourselves that the exact expression
for the survival probability with $\sigma\!=\!1$ is
\begin{eqnarray}
Q_{1}(t\!\mid\!x_{0})&=&1+\frac{3\sqrt{3}}{2\pi}\,\Theta\!\left(t-\frac{2\lvert x_{0}\rvert^{3}}
{9D}\right)\left(\frac{9Dt}{2\lvert x_{0}\rvert^{3}}-1\right)^{\frac{1}{3}}\nonumber \\
&\times&\bigg[\pFq{2}{1}\left(1,1,\frac{4}{3},1-\frac{9Dt}{2\lvert x_{0}\rvert^{3}}\right)
-\pFq{2}{1}\left(1,\frac{1}{3},\frac{4}{3},1-\frac{9Dt}{2\lvert x_{0}\rvert^{3}}\right)\bigg].
\label{eq_40}
\end{eqnarray}
In turn, Eqs.~(\ref{eq_C6}) and (\ref{eq_C7}) derived in Appendix~C allow us to Laplace
inverse the function in Eq.~(\ref{eq_39}) and eventually obtain the survival probability for
the parameter $\sigma\!=\!2$, which is as follows:
\begin{eqnarray}
Q_{2}(t\!\mid\!x_{0})&=&1+\frac{\Gamma^{2}\!\left(\frac{1}{4}\right)\lvert x_{0}\rvert}
{4\sqrt{2}\,\pi D^{1/4}}\,\Theta\!\left(t-\frac{\pi^{2}x_{0}^{4}}{16D}\right)
\left(t-\frac{\pi^{2}x_{0}^{4}}{16D}\right)^{-\frac{1}{4}}\nonumber \\
&\times&\bigg[\left(\frac{\pi^{2}x_{0}^{4}}{16Dt}\right)^{\frac{1}{4}}
-\pFq{2}{1}\left(-\frac{1}{4},\frac{5}{4},\frac{3}{4},1-\frac{16Dt}{\pi^{2}
x_{0}^{4}}\right)\bigg].
\label{eq_41}
\end{eqnarray}
In the last two formulae, the notation $\pFq{2}{1}\left(a,b,c,z\right)$ represents the
three-parameter Gaussian hypergeometric function~\cite{Slat1966}. The time course of the
survival probability described by Eq.~(\ref{eq_40}) for $\sigma\!=\!1$ is plotted in
Fig.~\ref{fig2}, whereas the corresponding time course of the survival probability given by
Eq.~(\ref{eq_41}) for $\sigma\!=\!2$ is shown in Fig.~\ref{fig3}. In both the cases the two
different distances from the initial position $x_{0}$ to the target point at the origin of the
semi-infinite interval have been chosen and the diffusion coefficient $D\!=\!1.0$ has been
assumed. We see that the dependence of the survival probability on time consists of two distinct
phases. For the first period of time its value constantly remains equal to unity, including the
initial condition $Q(0\!\mid\!x_{0})\!=\!1$ at $t\!=\!0$, and only in the second phase
monotonically decreases to reach zero at infinity. This phase appears after the front of the
PDF, assigned on a finite support outside of which it disappears (see Fig.~\ref{fig1}), has
managed to attain the target point for the first time. Prior this event, the probability of
finding the diffusing particle at that point amounts exactly zero. It is not a difficult task
to prove that these properties emanate directly from the expressions embodied by
Eqs.~(\ref{eq_40}) and (\ref{eq_41}).
\begin{figure}[t]
\centering
\includegraphics[scale=0.35]{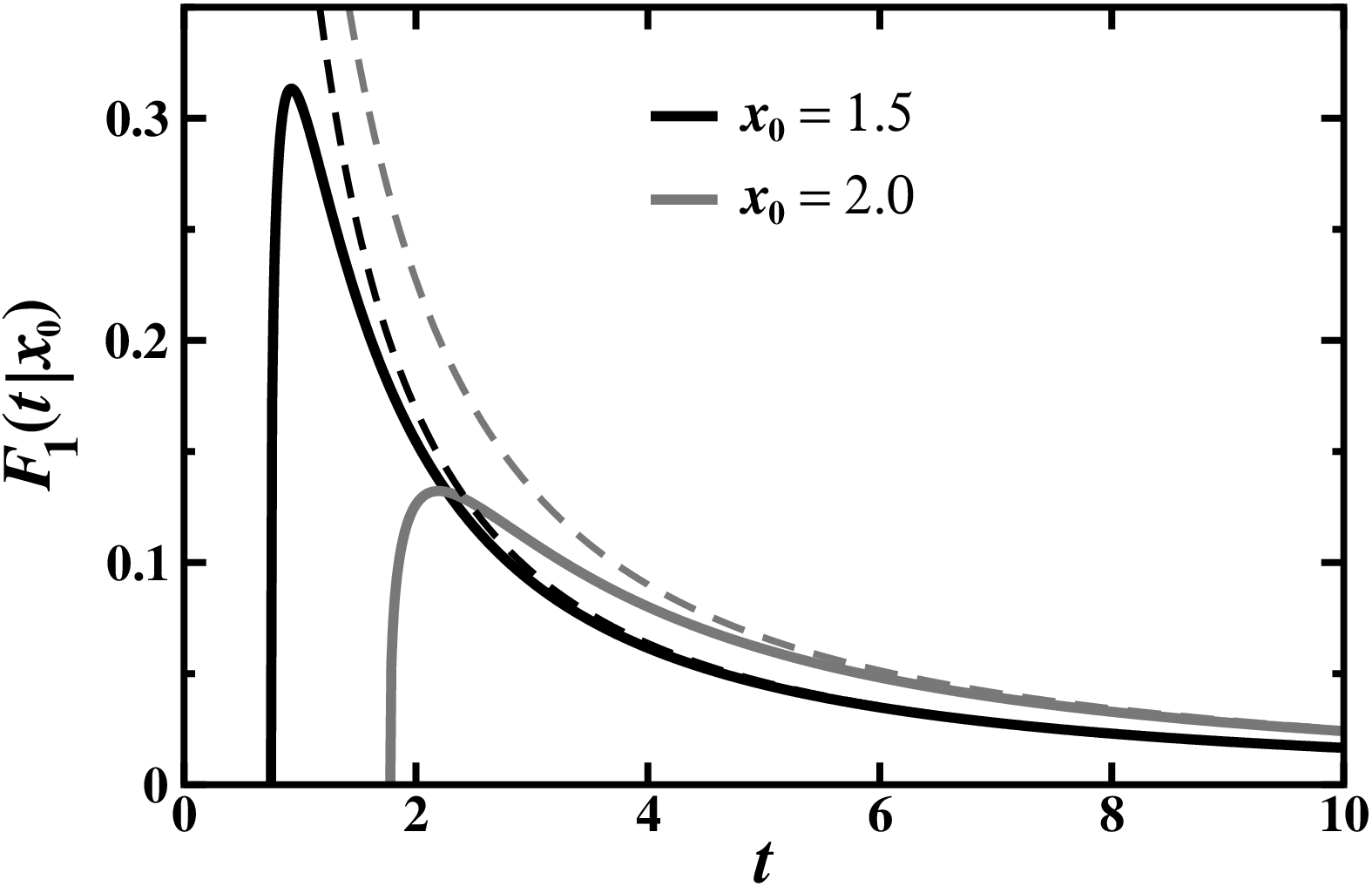}
\caption{Time course of the FPT distribution given by Eq.~(\ref{eq_42}) for the non-linear
diffusion with the parameter $\sigma\!=\!1$. The dashed lines refer to the asymptotic
representation of the FPT distribution shown in Eq.~(\ref{eq_44}). Here, two values of the
initial position $x_{0}$ relative to the target point located a the origin $x\!=\!0$ of the
semi-infinite interval have been chosen and the value of the diffusion coefficient $D\!=\!1.0$
has been assumed.}
\label{fig4}
\end{figure}
\begin{figure}[t]
\centering
\includegraphics[scale=0.35]{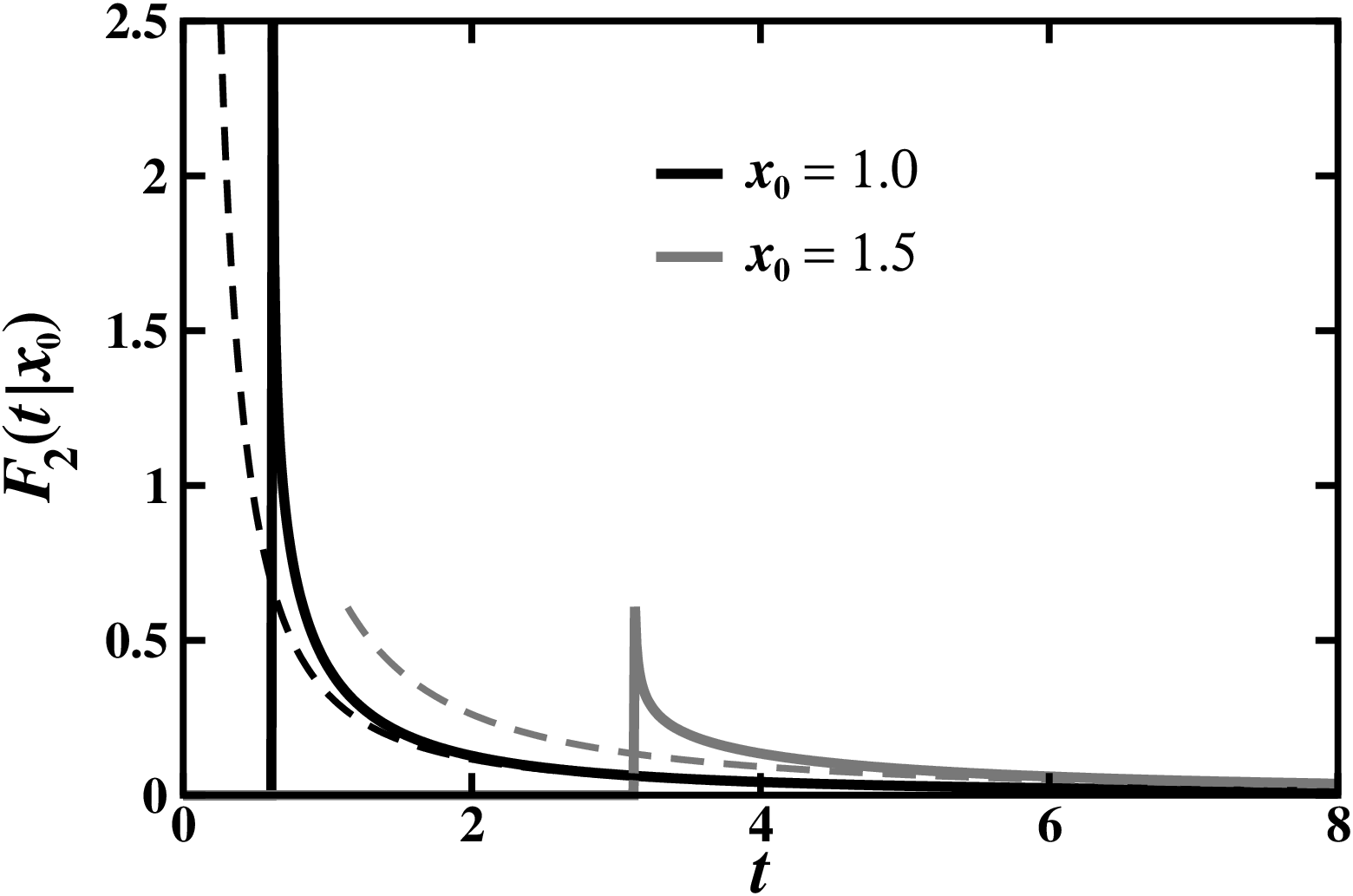}
\caption{Time course of the FPT distribution given by Eq.~(\ref{eq_43}) for the non-linear
diffusion with the parameter $\sigma\!=\!2$. The dashed lines refer to the asymptotic
representation of the FPT distribution shown in Eq.~(\ref{eq_45}). Here, two values of the
initial position $x_{0}$ relative to the target point located a the origin $x\!=\!0$ of the
semi-infinite interval have been chosen and the value of the diffusion coefficient $D\!=\!1.0$
has been assumed.}
\label{fig5}
\end{figure}

As noted in Sec.~\ref{sec_2} the first time derivative of the survival probability, preceded
with the negative sign, results in the FPT distribution (see Eq.~(\ref{eq_11})). Here, we
present the final formulae for this quantity characterising the first-passage statistics
of the non-linear diffusion with the parameter $\sigma\!=\!1$ and $2$. The method
leading to these results is detailed in Appendix~D. Thus, recalling Eq.~(\ref{eq_40}) and
appealing to Eq.~(\ref{eq_D13}), we show that the FPT distribution for $\sigma\!=\!1$ is
\begin{eqnarray}
F_{1}(t\!\mid\!x_{0})&=&\frac{\sqrt{3}}{2\pi t}\,\Theta\left(t-\frac{2\lvert x_{0}\rvert^{3}}
{9D}\right)\left(\frac{9Dt}{2\lvert x_{0}\rvert^{3}}-1\right)^{-\frac{2}{3}}
\bigg\{1+\frac{9Dt}{2\lvert x_{0}\rvert^{3}} \nonumber\\
&\times&\left[2\,\pFq{2}{1}\left(1,1,\frac{4}{3},1-\frac{9Dt}{2\lvert x_{0}\rvert^{3}}\right)
-3\,\pFq{2}{1}\left(1,2,\frac{4}{3},1-\frac{9Dt}{2\lvert x_{0}\rvert^{3}}\right)\right]
\bigg\},
\label{eq_42}
\end{eqnarray}
whereas the FPT distribution for $\sigma\!=\!2$ has the following form:
\begin{equation}
F_{2}(t\!\mid\!x_{0})=\frac{\Gamma^{2}\!\left(\frac{1}{4}\right)\lvert x_{0}\rvert}{8\sqrt{2}
\pi t(Dt)^{1/4}}\,\Theta\!\left(t-\frac{\pi^{2}x_{0}^{4}}{16D}\right)\left(\frac{16Dt}{\pi^{2}
x_{0}^{4}}-1\right)^{-\frac{1}{4}}.
\label{eq_43}
\end{equation}
The plots exposed in Figs.~\ref{fig4} and \ref{fig5} illustrate how the FPT distributions
depend on the time, correspondingly, for the parameter $\sigma\!=\!1$ and $2$. Here, we fixed
the diffusion coefficient $D\!=\!1.0$. All the profiles of these functions are shaped by
selection of various distances from the initial position $x_{0}$ of the diffusing particle to
the target point located at the origin of the semi-infinite interval. The dashed lines
correspond to the asymptotic representations of FPT distributions (see Eqs.~(\ref{eq_44}) and
(\ref{eq_45})). Also in this case, the two phases in the time course of these distribution
functions can be distinguished. Where the value of the survival probability amounts one, the
FPT distribution vanishes.

The FPT distribution evaluates the likelihood when the diffusing particle hits the
pre-specified target for the first time. Its first moment defines the mean time upon which
this target might be achieved. We have argued in Sec.~\ref{sec_31} that the MFPT to the
origin of the semi-infinite interval, in which the linear diffusion proceeds, is divergent.
Does the same regularity manifest itself in the case of the non-linear diffusion with the
parameter $\sigma\!=\!1$ and $2$? The normalization condition included in Eq.~(\ref{eq_15})
assures the particle will arrive at the target irrespective of the type of diffusive motion.
The FPT distributions given by Eqs.~(\ref{eq_42}) and (\ref{eq_43}) for the non-linear
diffusion are also normalized to unity. We can formally demonstrate this property for
$F_{1}(t\!\mid\!x_{0})$ by means of the two integrals:
$\int_{0}^{\infty}z^{-\mu-1}\pFq{2}{1}\left(\alpha,\beta,\gamma,-z\right)\mathrm{d}z\!=\!
\frac{\Gamma(\alpha+\mu)\Gamma(\beta+\mu)\Gamma(\gamma)\Gamma(-\mu)}{\Gamma(\alpha)\Gamma(\beta)
\Gamma(\gamma+\mu)}$, that holds for $\gamma\!\neq\!0,-1,-2,\dots$, $\mathrm{Re}\,\mu\!<\!0$,
$\mathrm{Re}(\alpha+\mu)\!>\!0$ and $\mathrm{Re}(\alpha+\mu)\!>\!0$, and
$\int_{0}^{\infty}z^{\mu-1}(z+1)^{-\nu}\mathrm{d}z\!=\!\mathrm{B}(\mu,\nu-\mu)\!=\!\frac{
\Gamma(\mu)\,\Gamma(\nu-\mu)}{\Gamma(\nu)}$, which is satisfied if $0\!<\!\mathrm{Re}\,\mu\!<\!
\mathrm{Re}\,\nu$. Only the second integral is needed to confirm the normalization of
$F_{2}(t\!\mid\!x_{0})$. The fact that both FPT distributions are normalized to unity implicates
that the diffusing particle is sure to reach the origin. Nevertheless, the MFPT turns out to be
infinite as in the case of the ordinary diffusion. We can verify this property inserting
Eqs.~(\ref{eq_42}) and (\ref{eq_43}) in Eq.~(\ref{eq_12}), or Eqs.~(\ref{eq_40}) and
(\ref{eq_41}) in Eq.~(\ref{eq_13}), and performing appropriate integration, or finally taking
the limit $s\!\to\!0$ in Eqs.~(\ref{eq_38}) and (\ref{eq_39}) following Eq.~(\ref{eq_33}).

The divergence of the MFPT for the non-linear diffusion with $\sigma\!=\!1$ and $2$ infers from
the long-time tails of the FPT distributions. With the asymptotic expansion of the Gaussian 
hypergeometric functions at hand, i.e. $\pFq{2}{1}\left(1,1,\frac{4}{3},1-a\,z\right)\!\propto\!
\frac{\sqrt{3}}{6\pi a}\Gamma\left(\frac{1}{3}\right)\Gamma\left(\frac{2}{3}\right)
\frac{\log(az)}{z}$ and $\pFq{2}{1}\left(1,2,\frac{4}{3},1-a\,z\right)\!\propto\!\frac{1}{3az}$,
where $a\!>\!0$ is required for both the cases, we show that
\begin{equation}
F_{1}(t\!\mid\!x_{0})\propto\frac{x_{0}^{2}}{9\,(\sqrt{6}D)^{2/3}\,t^{5/3}}
\left[1+\frac{\sqrt{3}}{\pi}\log\!\left(27\left(\frac{9Dt}{2\lvert
x_{0}\rvert^{3}}\right)^{2}\right)\right],
\label{eq_44}
\end{equation}
when $t\!\to\!\infty$, while the following long-time representation
\begin{equation}
F_{2}(t\!\mid\!x_{0})\propto\frac{\Gamma^{2}\!\left(\frac{1}{4}\right)x_{0}^{2}}
{16\sqrt{2\pi D}\,t^{3/2}}
\label{eq_45}
\end{equation}
straightforwardly emerges form Eq.~(\ref{eq_43}). Therefore, using Eq.~(\ref{eq_12}) along
with Eq.~(\ref{eq_44}) for $\sigma\!=\!1$, we indeed obtain that $\mathcal{T}_{1}(x_{0})\!=\!
\int_{0}^{\infty}t\,F_{1}(t\!\mid\!x_{0})\,\mathrm{d}t\!\sim\!\int_{0}^{\infty}t\times t^{-5/3}
\,\mathrm{d}t\!=\!\infty$. Here, we did not intentionally include the logarithmic correction
with the power-law factor $t^{-5/3}$.
\begin{figure}[t]
\centering
\includegraphics[scale=0.35]{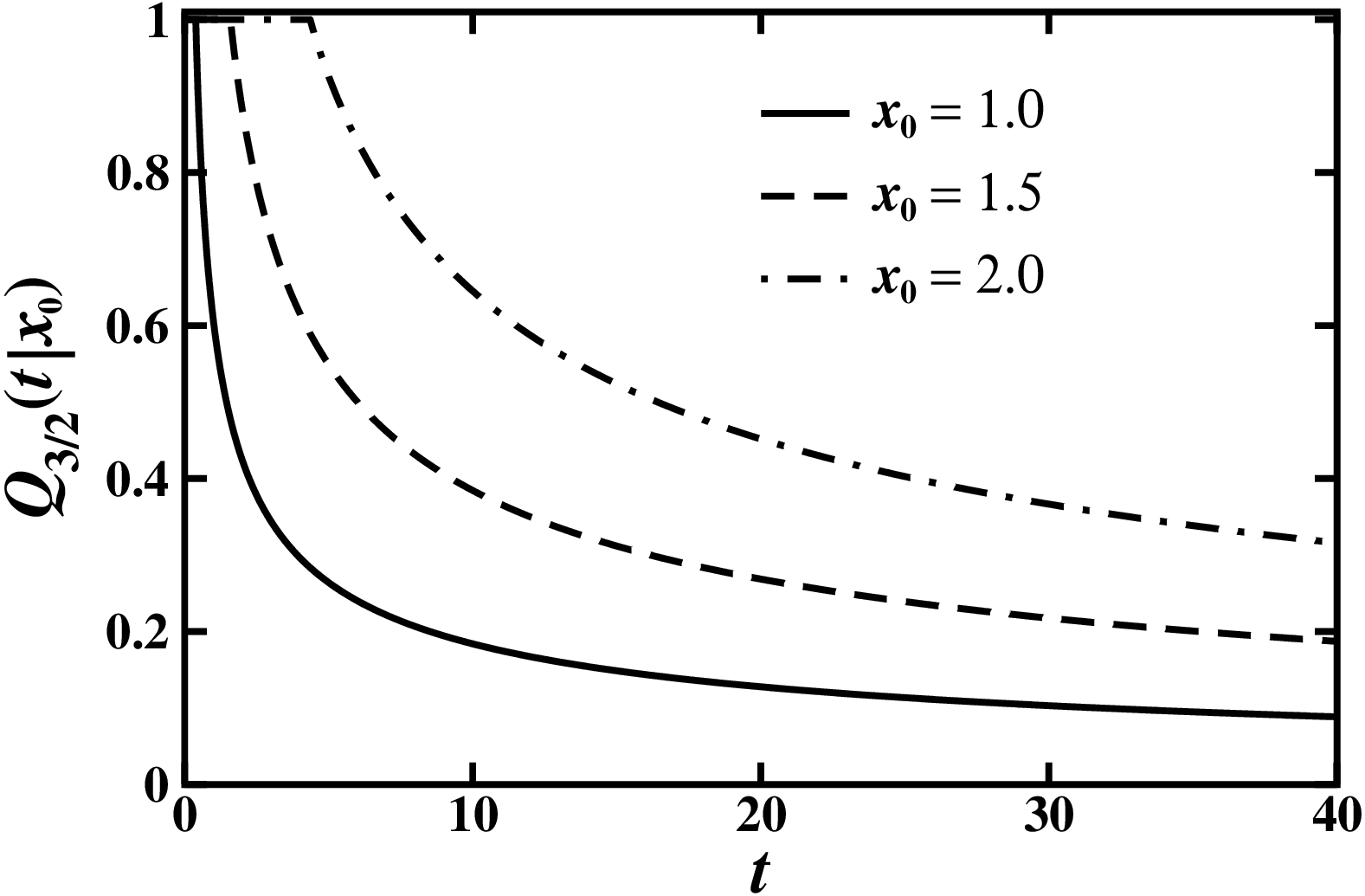}
\caption{Time course of the survival probability given by Eq.~(\ref{eq_49}) for the non-linear
diffusion with the parameter $\sigma\!=\!3/2$. Here, three values of the initial position $x_{0}$
relative to the target point located a the origin $x\!=\!0$ of the semi-infinite interval have
been chosen and the value of the diffusion coefficient $D\!=\!1.0$
has been assumed.}
\label{fig6}
\end{figure}
\begin{figure}[b]
\centering
\includegraphics[scale=0.3]{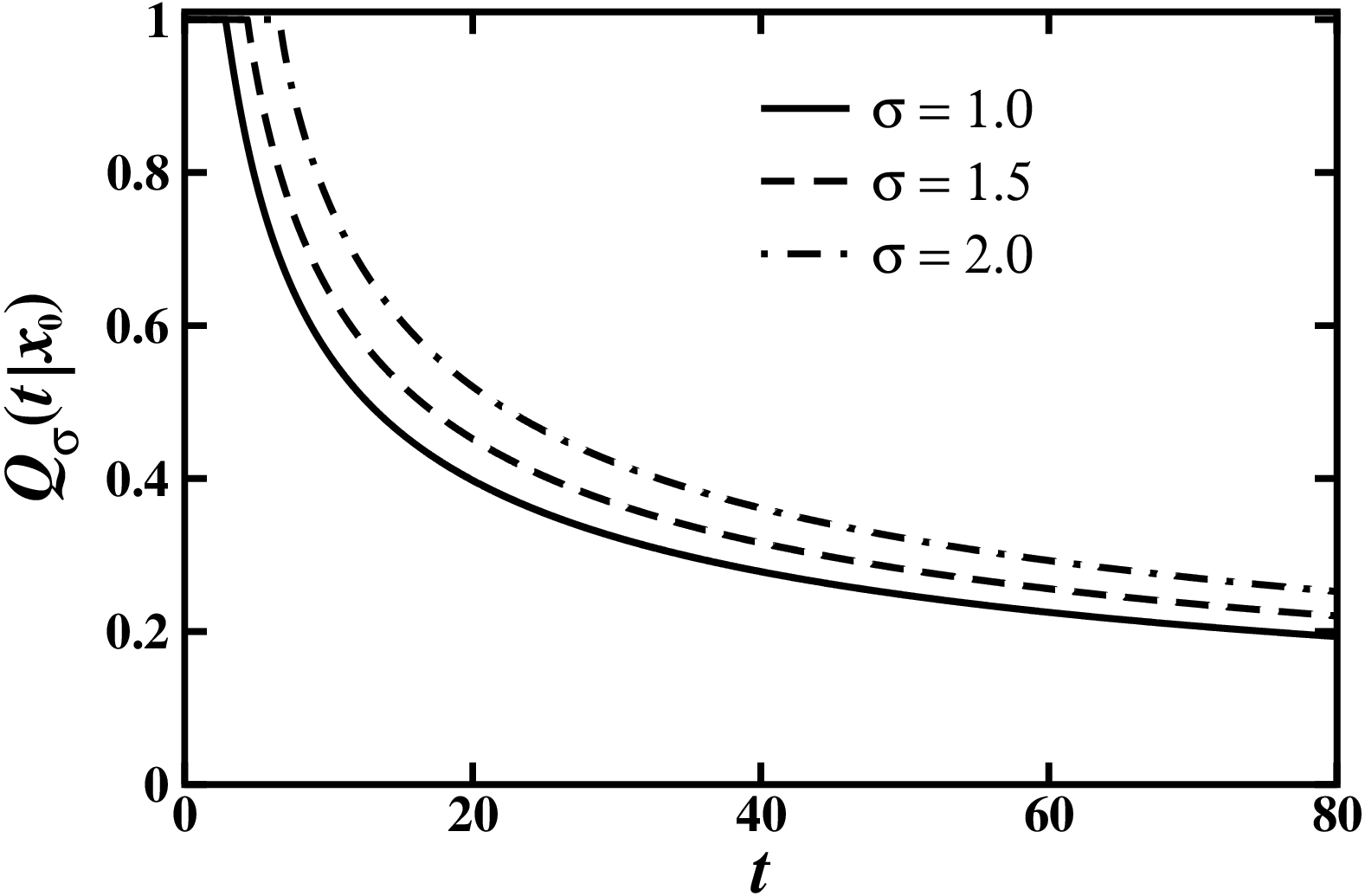}
\caption{Time course of the survival probability given by Eq.~(\ref{eq_49}) for the non-linear
diffusion with three exemplary values of the parameter $\sigma$. Here, the value of the initial
position $x_{0}\!=\!2$ relative to the target point located at the origin $x\!=\!0$ of the
semi-infinite interval has been chosen and the value of the diffusion coefficient $D\!=\!1.0$
has been assumed.}
\label{fig7}
\end{figure}
Similarly, we show taking advantage of Eq.~(\ref{eq_45}) that the MFPT $\mathcal{T}_{2}(x_{0})\!=\!\int_{0}^{\infty}t\,F_{2} (t\!\mid\!x_{0})\,
\mathrm{d}t\!\sim\!\int_{0}^{\infty}t\times t^{-3/2}\,\mathrm{d}t\!=\!\infty$ for
$\sigma\!=\!2$.

\subsubsection{\label{sec_4_2_2}Approximate results}

The duo of exact results derived in Sec.~\ref{sec_4_2_1} for the non-linear diffusion with the
diffusivity specified by the power-law exponent $\sigma\!=\!1$ and $2$ is exceptional.
This circumstance raises the natural question about remaining values of the parameter $\sigma$
in Eq.~(\ref{eq_2}). As aforementioned in Sec.~\ref{sec_4_2} the precise integration in
Eq.~(\ref{eq_35}) poses serious difficulties for arbitrary values of this parameter. To overcome
this problem, we will replace the integrand in Eq.~(\ref{eq_35}) with its Taylor expansion. Such
an approach is justified due to the following argumentation. Each PDF is by assumption a
non-negative and real (non-complex) quantity. Specifically, the power-law component of the PDF
in Eq.~\ref{eq_35}, i.e. $(1-z)^{\alpha}$ with $z:=[\eta(x,\sigma)/t]^{2/(\sigma+2)}$ and
$\alpha\!=\!1/\sigma$, satisfies this crucial requirement if and only if $\lvert z\rvert\!<\!1$.
Consequently, the corresponding inequality $\eta(x,\sigma)\!:=\!D^{-1}(\sqrt{\mathcal{B}
(\sigma)}\,\lvert x\rvert)^{\sigma+2}\!<\!t$ must be met, which allows us to approximate the
power-law function in Eq.~\ref{eq_35} by expanding it in the Taylor series, namely
$(1-z)^{\alpha}\!\approx\!1-\alpha z$. To complement the above reasoning, we would like to
emphasize that our preliminary studies confirmed the effectiveness of this approximation as
long as the exponent $\alpha\!<\!1$, what implicates the parameter $\sigma\!>\!1$. Consequently,
the approximate expression for the Laplace transform of the propagator from a position
$x'\!>\!0$ to the origin of the semi-infinite interval is
\begin{equation}
\tilde{p}(0,s\!\mid\!x')\simeq\frac{\mathcal{A}(\sigma)}{D^{\frac{1}{\sigma+2}}
s^{\frac{\sigma+1}{\sigma+2}}}\left[\Gamma\left(\frac{\sigma+1}{\sigma+2},s\,\eta(x',\sigma)
\right)-\frac{1}{\sigma}\left(s\,\eta(x',\sigma)\right)^{\frac{2}{\sigma+2}}
\Gamma\left(\frac{\sigma-1}{\sigma+2},s\,\eta(x',\sigma)\right)\right].
\label{eq_46}
\end{equation}

\begin{figure}[t]
\centering
\includegraphics[scale=0.35]{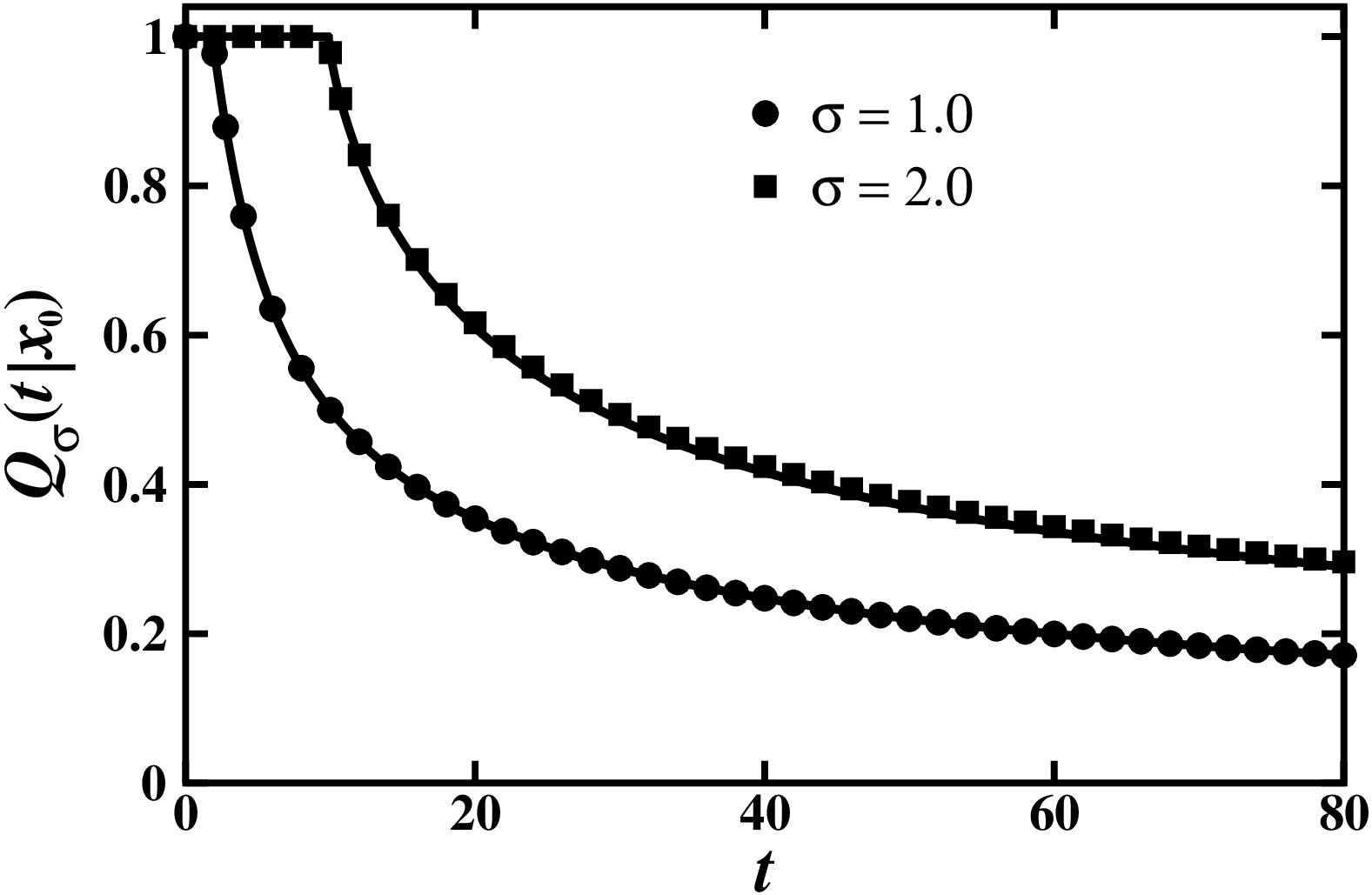}
\caption{The comparison of the approximate formula in Eq.~(\ref{eq_49}) (solid lines) to the
exact formulae in Eqs.~(\ref{eq_40}) (circles, $\sigma\!=\!1$) and (\ref{eq_41}) (squares,
$\sigma\!=\!2$) describing the time course of the survival probability for the non-linear
diffusion in the semi-infinite interval. Here, the value of the initial position $x_{0}\!=\!2$
relative to the target point located at the origin $x\!=\!0$ has been fixed and the value of
the diffusion coefficient $D\!=\!1.0$ has been assumed.}
\label{fig8}
\end{figure}
\noindent In order to obtain this result, we have utilized the following integral
$\int_{z}^{\infty}u^{\alpha-1}\mathrm{e}^{-u}\mathrm{d}u\!=\!\Gamma(\alpha,z)$, representing
the upper incomplete gamma function. Setting $x'\!=\!0$ causes that the auxiliary function
$\eta(0,\sigma)\!=\!0$ and the second term enclosed by the square bracket in Eq.~(\ref{eq_46})
disappears. But, the remaining upper incomplete gamma function turns into the Euler gamma
function, thus
\begin{equation}
\tilde{p}(0,s\!\mid\!0)\simeq\frac{\mathcal{A}(\sigma)}{D^{\frac{1}{\sigma+2}}
s^{\frac{\sigma+1}{\sigma+2}}}\,\Gamma\left(\frac{\sigma+1}{\sigma+2}\right).
\label{eq_47}
\end{equation}
A direct substitution of Eqs.~(\ref{eq_46}) and (\ref{eq_47}) in Eq.~(\ref{eq_32}) yields the
approximate formula for the Laplace transform of the survival probability
\begin{align}
\tilde{Q}_{\sigma}(s\!\mid\!x_{0})&\simeq\frac{1}{s}\bigg\{
1-\frac{1}{\Gamma\left(\frac{\sigma+1}{\sigma+2}\right)}
\bigg[\Gamma\left(\frac{\sigma+1}{\sigma+2},s\eta(x_{0},\sigma)\right)\nonumber\\
&-\frac{1}{\sigma}\left(s\eta(x_{0},\sigma)\right)^{\frac{2}{\sigma+2}}
\Gamma\left(\frac{\sigma-1}{\sigma+2},s\eta(x_{0},\sigma)\right)\bigg]
\bigg\},
\label{eq_48}
\end{align}
where now $\eta(x_{0},\sigma)\!=\!D^{-1}(\sqrt{\mathcal{B}(\sigma)}\,\lvert x_{0}\rvert)
^{\sigma+2}$ depends solely on the initial position $x_{0}$ of the diffusing particle.
Surprisingly, despite apparent complexity of this function, we can execute the inverse
Laplace transformation in order to find the survival probability depending on the time
variable. To this end, it is enough to refer to Eq.~(\ref{eq_B8}) derived in Appendix~B
and perform simple algebraic operations. The final result reads
\begin{eqnarray}
Q_{\sigma}(t\mid x_{0})&\simeq&1-\frac{\Theta[t-\eta(x_{0},\sigma)]}{\mathrm{B}\left(\frac{
\sigma+1}{\sigma+2},\frac{\sigma+3}{\sigma+2}\right)}\left(\frac{t}{\eta(x_{0},\sigma)}-1
\right)^{\frac{1}{\sigma+2}}\pFq{2}{1}\!\left(1,\frac{1}{\sigma+2},\frac{\sigma+3}{\sigma+2},
1-\frac{t}{\eta(x_{0},\sigma)}\right) \nonumber\\
&\times&\left[1-\frac{\pFq{2}{1}\!\left(1,\frac{3}{\sigma+2},\frac{\sigma+3}{\sigma+2},
1-\frac{t}{\eta(x_{0},\sigma)}\right)}{\pFq{2}{1}\!\left(1,\frac{1}{\sigma+2},\frac{\sigma+3}
{\sigma+2},1-\frac{t}{\eta(x_{0},\sigma)}\right)}\right]^{\frac{1}{\sigma}}.
\label{eq_49}
\end{eqnarray}
\begin{figure}[t]
\centering
\includegraphics[scale=0.35]{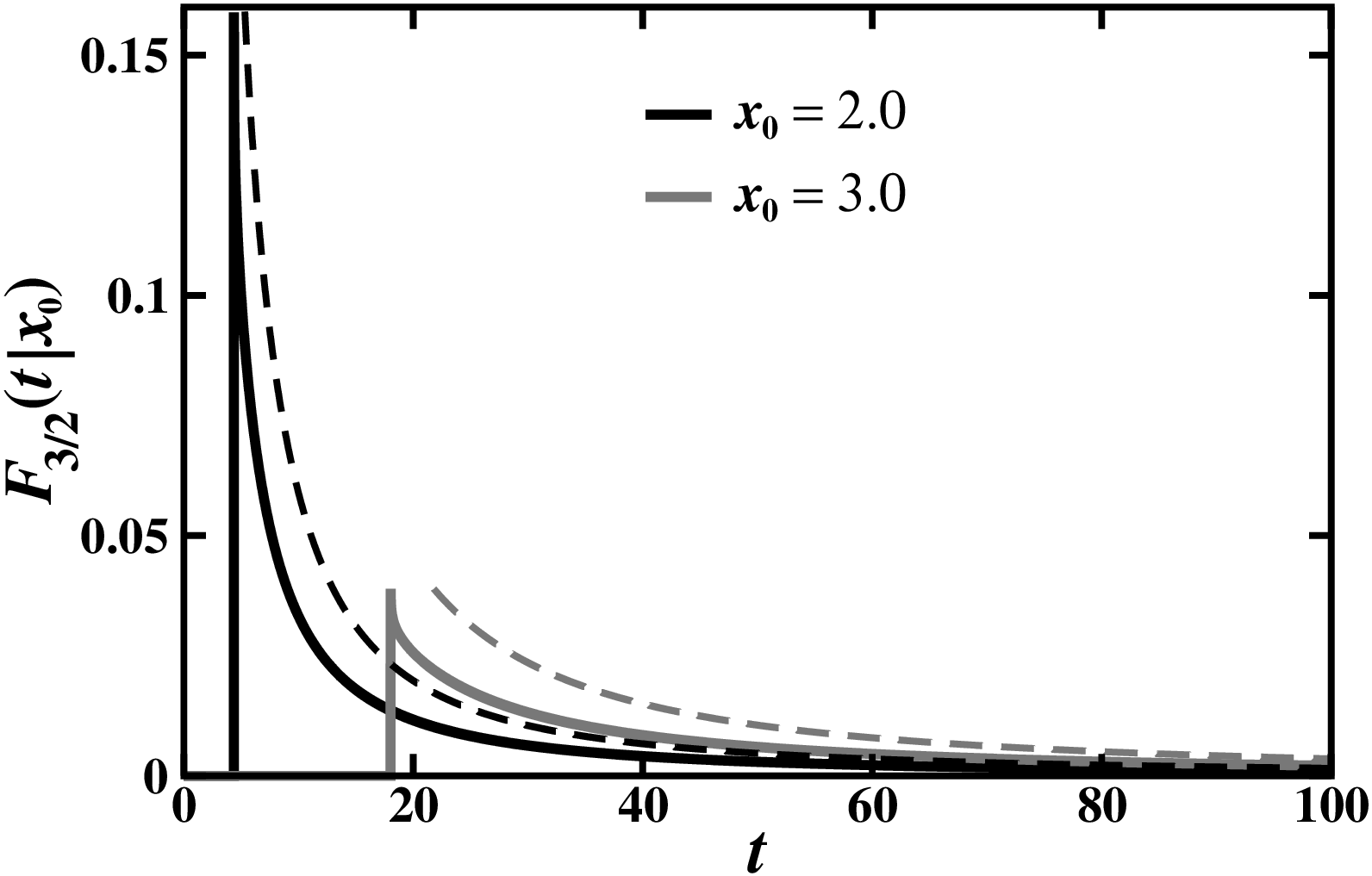}
\caption{Time course of the FPT distribution given by Eq.~(\ref{eq_50}) for the non-linear
diffusion with the parameter $\sigma\!=\!3/2$. The dashed lines refer to the asymptotic
representation of the FPT distribution shown in Eq.~(\ref{eq_52}). Here, two values of the
initial position $x_{0}$ relative to the target point located at the origin $x\!=\!0$ of the
semi-infinite interval have been chosen and the value of the diffusion coefficient $D\!=\!1.0$
has been assumed.}
\label{fig9}
\end{figure}

Now we examine whether this survival probability satisfies the two fundamental
properties which have been discussed in the previous subsection and re-visited in
Sec.~\ref{sec_3}. The first property corresponding to the initial condition
$Q(0\!\mid\!x_{0})\!=\!1$ is satisfied by Eq.~(\ref{eq_49}) due to the presence of the Heviside
unit step function. The second property concerning the monotonic decrease of the survival
probability in the long-time limit, i.e. $Q(t\!\mid\!x_{0})\!\to\!0$ if $t\!\to\!\infty$, is
easy to test considering the asymptotic expansions of the Gaussian hypergeometric functions
$\pFq{2}{1}\!\left(1,\frac{1}{\sigma+2},\frac{\sigma+3}{\sigma+2},1-z\right)\!\propto\!
z^{-1/(\sigma+2)}\;\Gamma\left(\frac{\sigma+1}{\sigma+2}\right)\Gamma\left(\frac{\sigma+3}
{\sigma+2}\right)$ and $\pFq{2}{1}\!\left(1,\frac{3}{\sigma+2},\frac{\sigma+3}{\sigma+2},1-z
\right)\!\propto\!z^{-3/(\sigma+2)}\;\Gamma\left(\frac{\sigma+3}{\sigma+2}\right)\Gamma
\left(\frac{\sigma-1}{\sigma+2}\right)/\,\Gamma\left(\frac{\sigma}{\sigma+2}\right)$ for
$z\!\to\!\infty$. Alternatively, we can take advantage of the well known limit theorems and
apply them to Eq.~(\ref{eq_48}). The first proposition applicable to the initial condition
states that if $t\!=\!0$, then $Q(0\!\mid\!x_{0})\!=\!\lim_{t\to0}Q(t\!\mid\!x_{0})\!=\!
\lim_{s\to\infty}s\,\tilde{Q}(s\!\mid\!x_{0})$, whereas the second proposition states that
$Q(\infty\!\mid\!x_{0})\!=\!\lim_{t\to\infty} Q(t\!\mid\!x_{0})\!=\!\lim_{s\to0}s\,\tilde{Q}
(s\!\mid\!x_{0})$~\cite{Schi1999}. In addition, it is suffice to note that $\lim_{z\to0}
\Gamma(\alpha,z)\!=\!\Gamma(\alpha)$ and the asymptotic representation of the upper incomplete
gamma function $\Gamma(\alpha,z)\!\propto\!z^{\alpha-1}\mathrm{e}^{-z}$ for $\lvert z\rvert
\!\to\!\infty$. Then, by virtue of the limit theorems applied to Eq.~(\ref{eq_48}), we
immediately conclude that $Q(0\!\mid\!x_{0})\!=\!1$ and $Q(t\!\mid\!x_{0})\!\to\!0$ for
$t\!\to\!\infty$. These two properties are reflected in Figs.~(\ref{eq_6}) and (\ref{eq_7}),
where the time courses of the survival probability given by Eq.~(\ref{eq_49}) are displayed,
respectively, for various distances between the initial position $x_{0}$ of the diffusing
particle and the target point settled at the origin $x\!=\!0$ of the semi-infinite interval,
and three disparate values of the parameter $\sigma$. Moreover, in Fig.~(\ref{fig8}), we
demonstrate a surprising compatibility of the survival probability described by
Eq.~(\ref{eq_49}) with the corresponding exact formulae derived for $\sigma\!=\!1$ and $2$ in
Eqs.~(\ref{eq_40}) and (\ref{eq_41}), respectively.
\begin{figure}[t]
\centering
\includegraphics[scale=0.35]{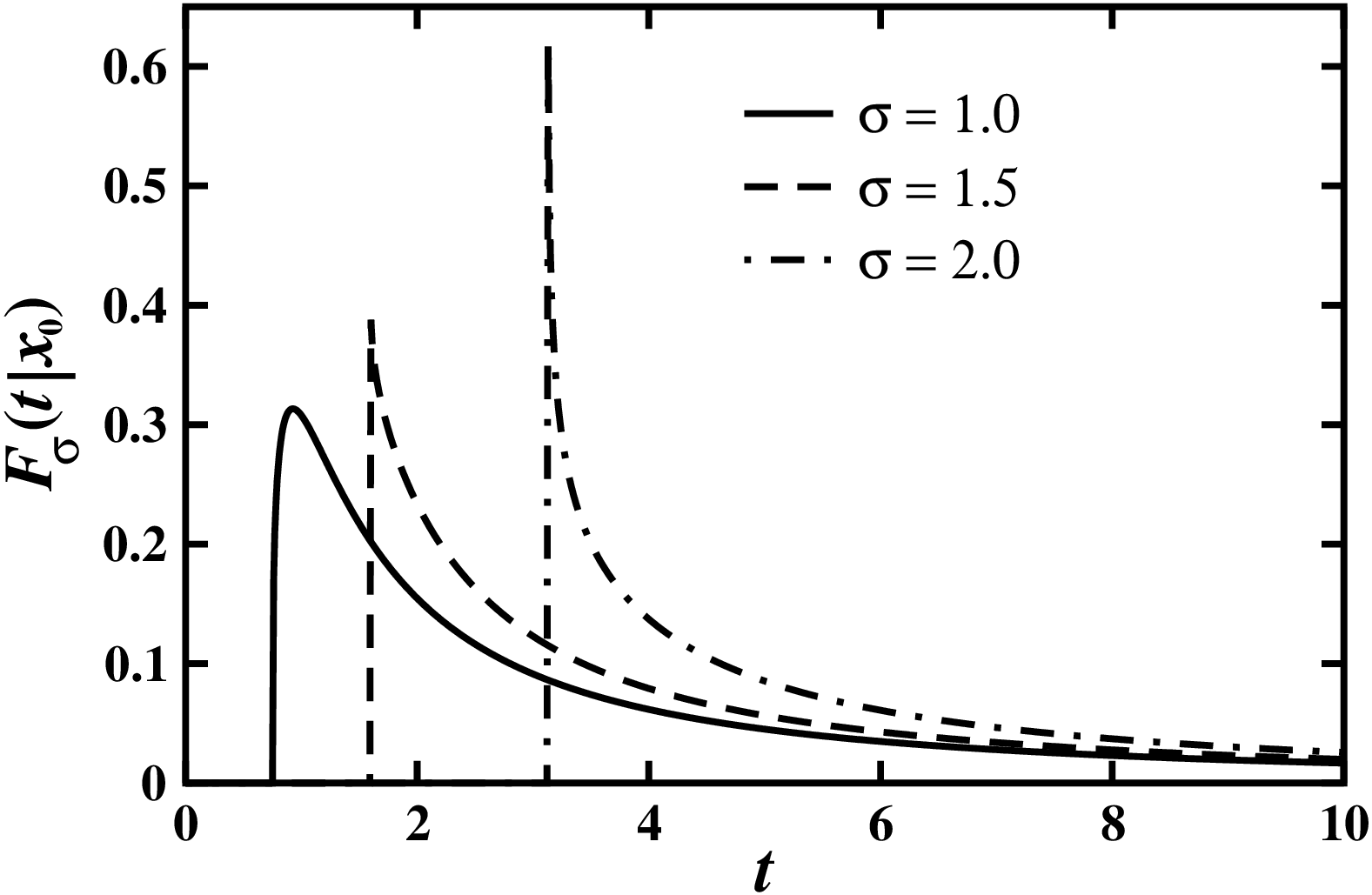}
\caption{Time course of the FPT distribution given by Eq.~(\ref{eq_50}) for the non-linear
diffusion with three exemplary values of the parameter $\sigma$. Here, the value of the initial
position $x_{0}\!=\!1.5$ relative to the target point located at the origin $x\!=\!0$ of the
semi-infinite interval has been chosen and the value of the diffusion coefficient $D\!=\!1.0$
has been assumed.}
\label{fig10}
\end{figure}

\begin{figure}[t]
\centering
\includegraphics[scale=0.35]{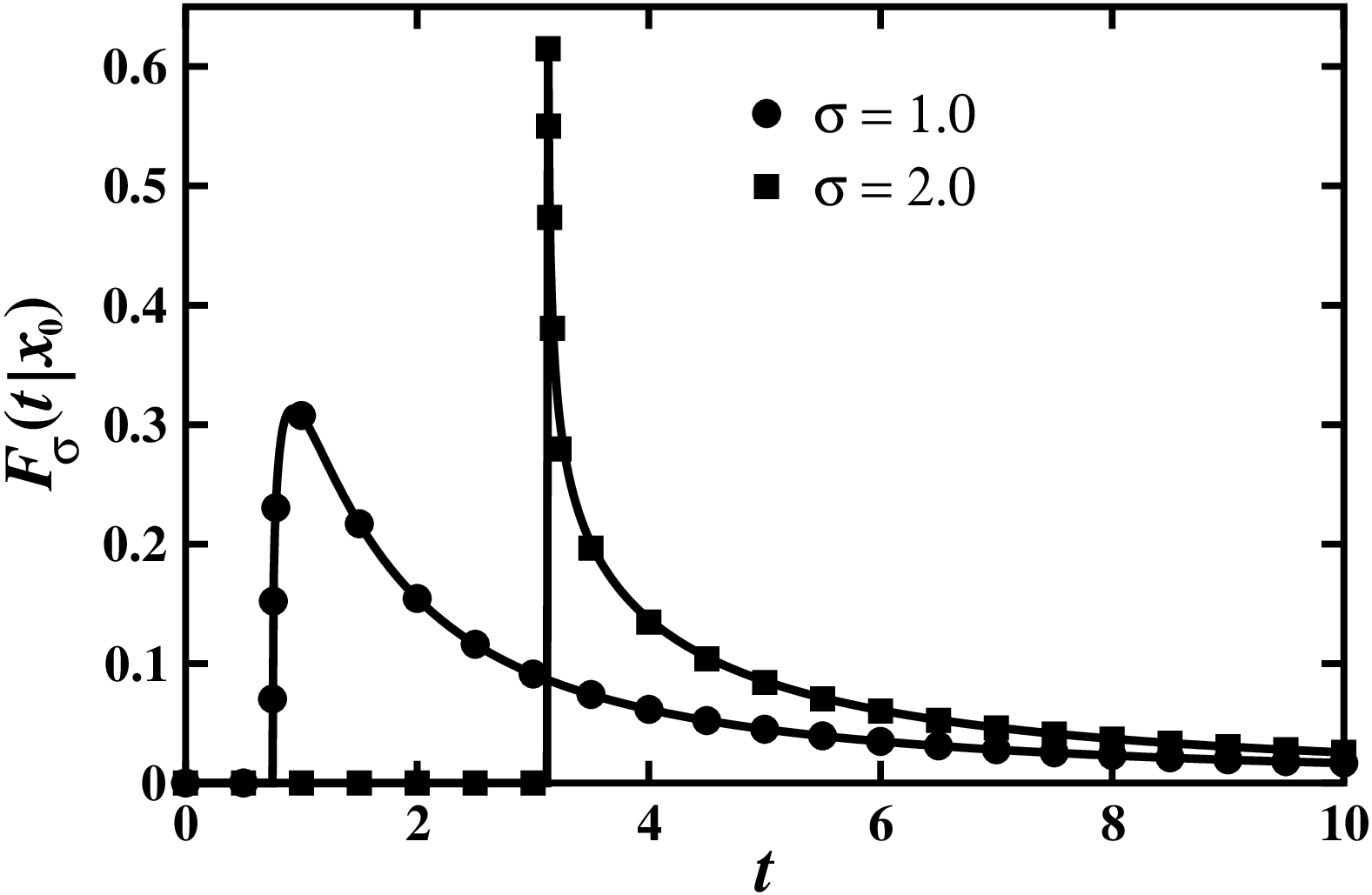}
\caption{The comparison of the approximate formula in Eq.~(\ref{eq_50}) (solid lines) to the
exact formulae in Eqs.~(\ref{eq_42}) (circles, $\sigma\!=\!1$) and (\ref{eq_43}) (squares,
$\sigma\!=\!2$) describing the time course of the FPT distribution for the non-linear
diffusion in the semi-infinite interval. Here, the value of the initial position $x_{0}\!=\!1.5$
relative to the target point located at the origin $x\!=\!0$ has been fixed and the value of
the diffusion coefficient $D\!=\!1.0$ has been assumed.}
\label{fig11}
\end{figure}

Given the survival probability in Eq.~(\ref{eq_49}), we are now ready to find the approximate
formula representing the FPT distribution for the non-linear diffusion in the semi-infinite
interval. Taking into account Eq.~(\ref{eq_11}) and determining the first time derivative of
the function shown in Eq.~(\ref{eq_49}), we obtain after quite long calculations that
\begin{eqnarray}
F_{\sigma}(t\!\mid\!x_{0})&\simeq&\frac{\Theta[t-\eta(x_{0},\sigma)]}{(\sigma+2)\,
\mathrm{B}\left(\frac{\sigma+1}{\sigma+2},\frac{\sigma+3}{\sigma+2}\right)
[\eta(x_{0},\sigma)]^{1/(\sigma+2)}\,[t-\eta(x_{0},\sigma)]^{(\sigma+1)/(\sigma+2)}}
\nonumber\\
&\times&\left[1-\frac{\pFq{2}{1}\!\left(1,\frac{3}{\sigma+2},\frac{\sigma+3}{\sigma+2},
1-\frac{t}{\eta(x_{0},\sigma)}\right)}{\pFq{2}{1}\!\left(1,\frac{1}{\sigma+2},\frac{\sigma+3}
{\sigma+2},1-\frac{t}{\eta(x_{0},\sigma)}\right)}\right]^{\frac{1}{\sigma}}
\Biggl\{\pFq{2}{1}\!\left(1,\frac{\sigma+3}{\sigma+2},\frac{\sigma+3}{\sigma+2},1-\frac{t}
{\eta(x_{0},\sigma)}\right) \nonumber\\
&+&\frac{\mathcal{F}(t,x_{0},\sigma)}{\sigma\left[\pFq{2}{1}\!\left(1,\frac{1}{\sigma+2},
\frac{\sigma+3}{\sigma+2},1-\frac{t}{\eta(x_{0},\sigma)}\right)-\pFq{2}{1}\!\left(1,\frac{3}
{\sigma+2},\frac{\sigma+3}{\sigma+2},1-\frac{t}{\eta(x_{0},\sigma)}\right)\right]}\Biggl\},
\label{eq_50}
\end{eqnarray}
where
\begin{eqnarray}
\mathcal{F}(t,x_{0},\sigma)&=&\pFq{2}{1}\!\left(1,\frac{3}{\sigma+2},\frac{\sigma+3}{\sigma+2},
1-\frac{t}{\eta(x_{0},\sigma)}\right)
\pFq{2}{1}\!\left(1,\frac{\sigma+3}{\sigma+2},\frac{\sigma+3}{\sigma+2},
1-\frac{t}{\eta(x_{0},\sigma)}\right) \nonumber\\
&+&2\,\pFq{2}{1}\!\left(1,\frac{1}{\sigma+2},\frac{\sigma+3}{\sigma+2},
1-\frac{t}{\eta(x_{0},\sigma)}\right)
\pFq{2}{1}\!\left(1,\frac{3}{\sigma+2},\frac{\sigma+3}{\sigma+2},
1-\frac{t}{\eta(x_{0},\sigma)}\right) \nonumber\\
&-&3\,\pFq{2}{1}\!\left(1,\frac{1}{\sigma+2},\frac{\sigma+3}{\sigma+2},
1-\frac{t}{\eta(x_{0},\sigma)}\right)
\pFq{2}{1}\!\left(1,\frac{\sigma+5}{\sigma+2},\frac{\sigma+3}{\sigma+2},
1-\frac{t}{\eta(x_{0},\sigma)}\right) \nonumber\\
\label{eq_51}
\end{eqnarray}
and $\eta(x_{0},\sigma)\!=\!D^{-1}(\sqrt{\mathcal{B}(\sigma)}\,\lvert x_{0}\rvert)^{\sigma+2}$.
Fig.~\ref{fig9} displays the time course of the FPT distribution with the parameter $\sigma\!=\!
3/2$ for two different initial positions $x_{0}\!=\!2.0$ and $3.0$ relative to the target point
established at the origin of the semi-infinite interval, whereas Fig.~\ref{fig10} shows the
same behavior from the point of view of three different values of the parameter $\sigma$ and
the fixed value of the initial position $x_{0}\!=\!1.5$. In Fig.~\ref{fig11} we present the
excellent conformity of the approximate expression for the FPT distribution in Eq.~(\ref{eq_50})
to the exact formulae given by Eqs.~(\ref{eq_42}) and (\ref{eq_43}), correspondingly, for the
parameter $\sigma\!=\!1$ and $2$.
\begin{figure}[t]
\centering
\includegraphics[scale=0.35]{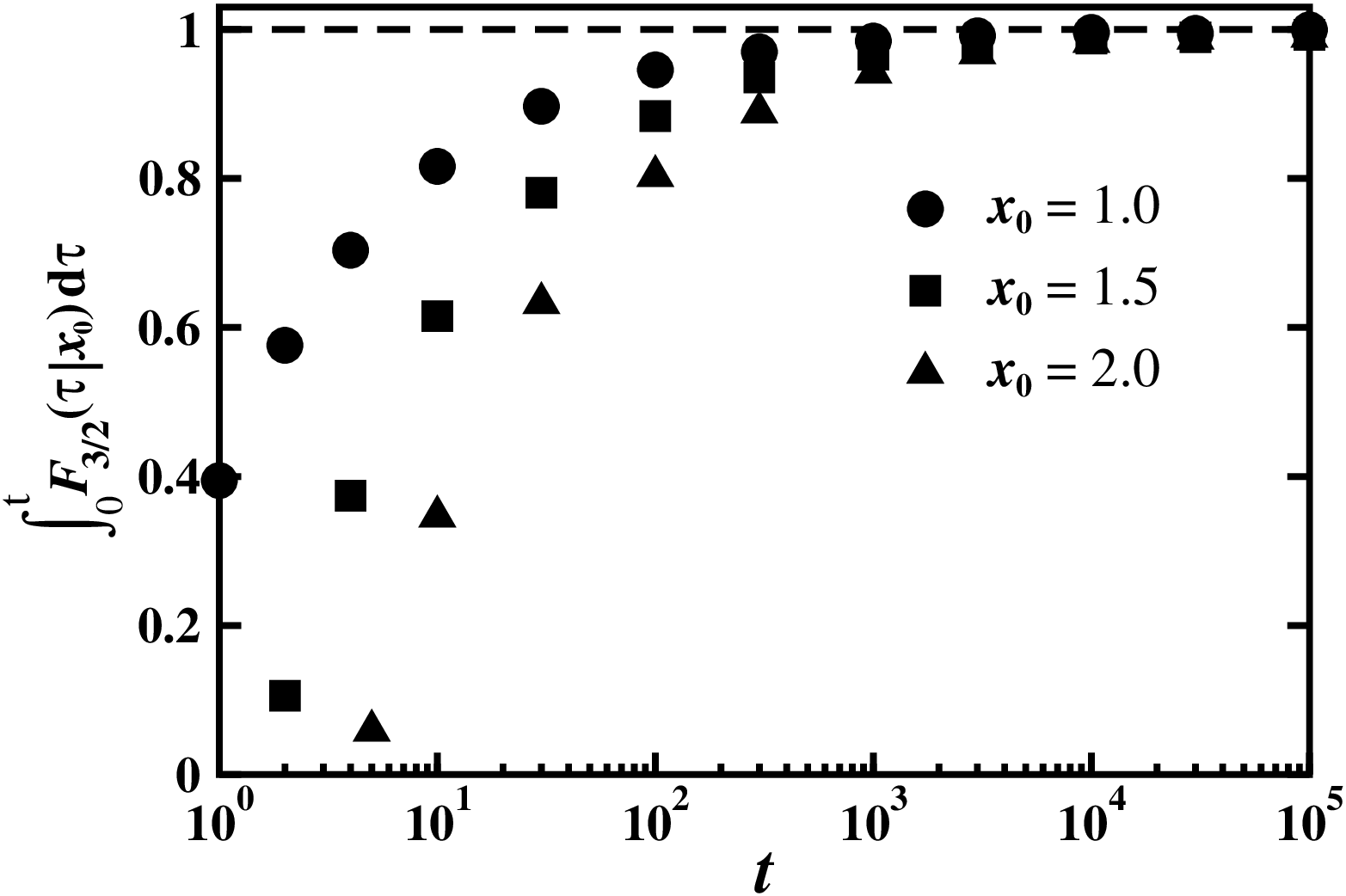}
\caption{Numerical test of the normalization condition $\int_{0}^{\infty}F_{\sigma}
(\tau\!\mid\!x_{0})\mathrm{d}\tau\!=\!1$ for the FPT distribution in Eq.~(\ref{eq_50}) with
the chosen parameter $\sigma\!=\!3/2$. Here, three values of the initial position $x_{0}$
relative to the target point located at the origin $x\!=\!0$ of the semi-infinite interval
has been fixed and the value of the diffusion coefficient $D\!=\!1.0$ has been assumed.}
\label{fig12}
\end{figure}

The analysis conducted in Sec.~\ref{sec_4_2_1} has shown that the FPT distributions for
the parameter $\sigma\!=\!1$ and $2$ are normalized to unity. This means that the diffusing
particle will definitely reach the pre-described target, although the MFPT needed to complete
this process is infinite. The same conclusion emerges from the formula in Eq.~(\ref{eq_50}),
but a direct integration of this function is rather too hard. Instead, we display in
Fig.~(\ref{fig12}) how the time integral of $F_{\sigma}(t\!\mid\!x_{0})$ converges to unity
with increasing range of numerical integration performed with respect to time for the parameter
$\sigma\!=\!3/2$, three fixed values of the initial position $x_{0}\!=\!1.0$, $1.5$ and $2.0$,
and the diffusion coefficient $D\!=\!1$. In turn, utilizing the asymptotic expansion of the FPT
distribution given by Eq.~(\ref{eq_50}), we can show a divergence of the MFPT. The asymptotic
formula is of the following form:
\begin{equation}
F_{\sigma}(t\!\mid\!x_{0})\propto\frac{2\sin\left(\frac{2\pi}{\sigma+2}\right)}
{\pi\sigma(\sigma+2)\,t}\,\mathrm{B}\!\left(\frac{\sigma-1}{\sigma+2},\frac{2}{\sigma+2}\right)
\!\left(\frac{\eta(x_{0},\sigma)}{t}\right)^{\frac{2}{\sigma+2}}.
\label{eq_52}
\end{equation}
The derivation of the above formula is detailed in Appendix~E. The divergence of the
MFPT is then due to the long-time tail of the FPT distribution. In essence, taking advantage
of Eq.~(\ref{eq_12}) we obtain that $\mathcal{T}_{\sigma}(x_{0})\!=\!\int_{0}^{\infty}t\,
F_{\sigma}(t\!\mid\!x_{0})\,\mathrm{d}t\!\sim\!\int_{0}^{\infty}t\times t^{-(\sigma+4)/
(\sigma+2)}\,\mathrm{d}t\!=\!\infty$ for $\sigma\!>\!0$. Again, this property is consistent
with the divergence of the MFPT for the ordinary diffusion in a semi-infinite interval
terminated by the totally absorbing wall~\cite{Redn2001}.

\section{\label{sec_5}Mean first-passage time for non-linear diffusion in harmonic potential}

We have shown in the previous section that the non-linear diffusion equation excludes the
finiteness of the MFPT to a target point located somewhere in an unbounded space. Such a property
is a part of a more general rule stating that any diffusive motion occurring in the unlimited
area of the space makes the MFPT infinite. However, this regularity changes if the diffusing
particle moves inside a bounded domain or in a confining potential. Here, we briefly explore the
latter scenario and consider the simple version of the non-linear diffusion in the harmonic
potential. A more substantial analysis of this issue represented by the following equation
\begin{equation}
\frac{\partial}{\partial t}p(x,t\!\mid\!x_{0})=\frac{\partial}{\partial x}
\left(\frac{\partial V(x)}{\partial x}p(x,t\!\mid\!x_{0})+
D\,p^{\sigma}(x,t\!\mid\!x_{0})\frac{\partial}{\partial x}p(x,t\!\mid\!x_{0})\right)
\label{eq_53}
\end{equation}
and extended to the other types of external potentials $V(x)$ will be a subject of an intense
research in the future.

Henceforth, we will study the diffusion equation of the harmonically trapped particle, whose
non-linearity is determined by the peculiar value of the parameter $\sigma\!=\!1$. Without loss
of generality, we assume the harmonic potential $V(x)\!=\!\frac{1}{2}\alpha x^{2}$ with the
certain stiffness $\alpha$ has the minimum at $x\!=\!0$. Due to these particular assumptions,
Eq.~(\ref{eq_53}) takes the form
\begin{equation}
\frac{\partial}{\partial t}p(x,t\!\mid\!x_{0})=\frac{\partial}{\partial x}
\left(\alpha x\,p(x,t\!\mid\!x_{0})+
D\,p(x,t\!\mid\!x_{0})\frac{\partial}{\partial x}p(x,t\!\mid\!x_{0})\right).
\label{eq_54}
\end{equation}
It is now convenient to substitute in the above differential equation $p(x,t\!\mid\!x_{0})
\!=\!\mathrm{e}^{\alpha t}q(z,\tau\!\mid\!z_{0})$, where $z\!=\!x\,\mathrm{e}^{\alpha t}$
and $\tau\!=\!\frac{1}{3\alpha}\left(\mathrm{e}^{3\alpha t}-1\right)$ for $t\!>\!0$, while
$z_{0}\!=\!x_{0}$ and $\tau\!=\!0$ for $t\!=\!0$. In this manner, we can transform
Eq.~(\ref{eq_54}) to much simpler form
\begin{equation}
\frac{\partial}{\partial\tau}q(z,\tau\!\mid\!z_{0})=D\frac{\partial}{\partial z}\left(
q(z,\tau\!\mid\!z_{0})\frac{\partial}{\partial z}q(z,\tau\!\mid\!z_{0})\right).
\label{eq_55}
\end{equation}
This equation is exactly the same as Eq.~(\ref{eq_3}) if the parameter $\sigma\!=\!1$. Let
us recall that its solution undergoing the normalization condition is manifested by
Eq.~(\ref{eq_6}) along with Eqs.~(\ref{eq_7}) and (\ref{eq_8}). Therefore, inserting there
$\sigma\!=\!1$, we readily have from Eq.~(\ref{eq_6}) that
\begin{equation}
q(z,\tau\!\mid\!z_{0})=\left(\frac{3}{32}\right)^{\frac{1}{3}}\left[\frac{1}{(D\tau)^{1/3}}
-\left(\frac{2}{9}\right)^{\frac{2}{3}}\frac{(z-z_{0})^{2}}{D\tau}\right]
\label{eq_56}
\end{equation}
is the solution of Eq.~(\ref{eq_55}). So, the exact solution of the original Eq.~(\ref{eq_54})
reads
\begin{equation}
p(x,t\!\mid\!x_{0})=\left(\frac{3}{32}\right)^{\frac{1}{3}}\mathrm{e}^{\alpha t}\left[
\left(\frac{3\alpha}{D}\right)^{\frac{1}{3}}\frac{1}{\left(\mathrm{e}^{3\alpha t}-1\right)
^{1/3}}-\left(\frac{2}{9}\right)^{\frac{2}{3}}\frac{3\alpha}{D}\frac{\left(x\mathrm{e}^{
\alpha t}-x_{0}\right)^{2}}{\mathrm{e}^{3\alpha t}-1}\right].
\label{eq_57}
\end{equation}
We can easy check this by a direct substitution of the above function in the partial
differential equation. From now on, we restrict our considerations to the non-linear diffusion
starting from the initial position at $x_{0}\!>\!0$ and progressing towards the target placed
in the minimum $x\!=\!0$ of the harmonic potential. In this particular case the PDF in
Eq.~(\ref{eq_57}) takes the simpler form
\begin{equation}
p(0,t\!\mid\!x_{0})=\left(\frac{3}{32}\right)^{\frac{1}{3}}\left[
\left(\frac{3\alpha}{D}\right)^{\frac{1}{3}}\frac{\mathrm{e}^{\alpha t}}
{\left(\mathrm{e}^{3\alpha t}-1\right) ^{1/3}}-\left(\frac{2}{9}\right)^{\frac{2}{3}}
\frac{3\alpha}{D}\frac{x_{0}^{2}\,\mathrm{e}^{\alpha t}}{\mathrm{e}^{3\alpha t}-1}\right],
\label{eq_58}
\end{equation}
which allows us to figure out the analytical expression for the MFPT downward the harmonic
potential.

The PDF must by definition be a non-negative quantity and this property in the case of
Eq.~(\ref{eq_58}) will be met if $t\!\geqslant\!\frac{1}{3\alpha}\log\left(1+\frac{2\alpha
\lvert x_{0}\rvert^{3}}{3D}\right)$. For this reason, the Laplace transformation of the PDF
is given by
\begin{equation}
\tilde{p}(0,s\!\mid\!x_{0})=\int_{0}^{\infty}\Theta\!\left[t-\frac{1}{3\alpha}
\log\left(1+\frac{2\alpha\lvert x_{0}\rvert^{3}}{3D}\right)\right]p(0,t\!\mid\!x_{0})
\mathrm{e}^{-st}\mathrm{d}t,
\label{eq_59}
\end{equation}
where $\Theta(z)$ denotes the Heviside unit step function. Plugging Eq.~(\ref{eq_58}) into
the above formula and taking advantage of the integral $\int_{u}^{\infty}\frac{z^{\mu-1}}
{(1+\alpha\,z)^{\nu}}\mathrm{d}z\!=\!\frac{u^{\mu-\nu}}{\alpha^{\nu}(\nu-\mu)}\,
\pFq{2}{1}\left(\nu,\nu-\mu,\nu-\mu+1,-\frac{1}{\alpha\,u}\right)$, which proceeds if
$\mathrm{Re}\,\nu\!>\mathrm{Re}\,\mu$, we obtain after straightforward calculations that
\begin{align}
\tilde{p}(0,s\!\mid\!x_{0})&=\left(\frac{3}{32}\right)^{\frac{1}{3}}\bigg[
\!\left(\frac{3\alpha}{D}\right)^{\frac{1}{3}}\frac{1}{s}\left(\frac{3D}{2\alpha\lvert x_{0}
\rvert^{3}}\right)^{\frac{s}{3\alpha}}\pFq{2}{1}\!\left(\frac{s}{3\alpha},\frac{s}{3\alpha}+
\frac{2}{3},\frac{s}{3\alpha}+1,-\frac{3D}{2\alpha\lvert x_{0}\rvert^{3}}\right) \nonumber\\
&-\left(\frac{2}{9}\right)^{\frac{2}{3}}\frac{3\alpha}{D}\frac{x_{0}^{2}}{s+2\alpha}\!
\left(\frac{3D}{2\alpha\lvert x_{0}\rvert^{3}}\right)^{\frac{s}{3\alpha}+\frac{2}{3}}
\pFq{2}{1}\!\left(\frac{s}{3\alpha}+\frac{2}{3},\frac{s}{3\alpha}+\frac{2}{3},\frac{s}
{3\alpha}+\frac{5}{3},-\frac{3D}{2\alpha\lvert x_{0}\rvert^{3}}\right)\!\bigg].
\label{eq_60}
\end{align}
On the other hand, setting above $x_{0}\!=\!0$ and utilizing the Laplace transformation
$\mathcal{L}\left[\left(1-\mathrm{e}^{-at}\right)^{\nu}\right]\\\!=\!a^{-1}\mathrm{B}
(\nu+1,,a^{-1}s)$, which is valid provided that $\mathrm{Re}\,a\!>\!0$ and $\nu\!>\!-1$,
we have
\begin{equation}
\tilde{p}(0,s\!\mid\!0)=\left(\frac{3}{32}\right)^{\frac{1}{3}}
\left(\frac{3\alpha}{D}\right)^{\frac{1}{3}}\frac{
\Gamma\left(\frac{2}{3}\right)\Gamma\left(\frac{s}{3\alpha}\right)}{3\alpha\,\Gamma\left(
\frac{s}{3\alpha}+\frac{2}{3}\right)},
\label{eq_61}
\end{equation}
where the Euler beta function $\mathrm{B}(x,y)\!=\!\frac{\Gamma(x)\Gamma(y)}{\Gamma(x+y)}$
has been simultaneously used. By virtue of Eq.~(\ref{eq_32}), the Laplace transform of the
survival probability
\begin{equation}
\tilde{Q}(s\!\mid\!x_{0})=\frac{\tilde{p}(0,s\!\mid\!0)-\tilde{p}(0,s\!\mid\!x_{0})}
{s\,\tilde{p}(0,s\!\mid\!0)}.
\label{eq_62}
\end{equation}
We can now combine Eqs.~(\ref{eq_60}) and (\ref{eq_61}) in order to insert them in the above
equation. This step leads to the following result
\begin{eqnarray}
\tilde{Q}(s\!\mid\!x_{0})&=&\frac{\Gamma\left(\frac{s}{3\alpha}+\frac{2}{3}\right)}
{\Gamma\left(\frac{2}{3}\right)\Gamma\left(\frac{s}{3\alpha}+1\right)}
\bigg[\frac{\Gamma\left(\frac{2}{3}\right)\Gamma\left(\frac{s}{3\alpha}+1\right)}{s\,
\Gamma\left(\frac{s}{3\alpha}+\frac{2}{3}\right)}-\frac{1}{s}\left(\frac{3D}{2\alpha\lvert
x_{0}\rvert^{3}}\right)^{s/3\alpha} \nonumber\\
&\times&\pFq{2}{1}\left(\frac{s}{3\alpha},\frac{s}{3\alpha}+
\frac{2}{3},\frac{s}{3\alpha}+1,-\frac{3D}{2\alpha\lvert x_{0}\rvert^{3}}\right)+
\left(\frac{2\alpha}{3D}\right)^{\frac{2}{3}}\frac{x_{0}^{2}}{s+2\alpha} \nonumber\\
&\times&\left(\frac{3D}{2\alpha\lvert x_{0}\rvert^{3}}\right)^{s/3\alpha+2/3}
\pFq{2}{1}\left(\frac{s}{3\alpha}+\frac{2}{3},\frac{s}{3\alpha}+
\frac{2}{3},\frac{s}{3\alpha}+\frac{5}{3},-\frac{3D}{2\alpha\lvert x_{0}\rvert^{3}}\right)
\bigg].
\label{eq_63}
\end{eqnarray}
The recipe embodied in Eq.~(\ref{eq_33}) allows us to find the MFPT directly from
Eq.~(\ref{eq_63}). For this purpose, it is enough to determine the limit of the Laplace transform
of the survival probability for $s\!\to\!0$. However, this operation yields the indeterminate
form $\infty-\infty$ and in consequence the need to apply L'Hospital's rule. Converting the
former indeterminate form to $\frac{0}{0}$ and using the L'Hospital rule twice, we definitively
show that
\begin{eqnarray}
\lim_{s\to0}&&\hspace{-5mm}\bigg[\frac{\Gamma\left(\frac{2}{3}\right)\Gamma\left(\frac{s}
{3\alpha}+1\right)}{s\,\Gamma\left(\frac{s}{3\alpha}+\frac{2}{3}\right)}-\frac{z^{s/3\alpha}}
{s}\,\pFq{2}{1}\left(\frac{s}{3\alpha},\frac{s}{3\alpha}+\frac{2}{3},\frac{s}{3\alpha}+1,-z\right)
\bigg] \nonumber\\
&=&-\frac{1}{3\alpha}\bigg[\frac{\partial}{\partial a}\,
\pFq{2}{1}\left(a,\frac{2}{3},1,-z\right)\!\bigg\rvert_{a=0}+\log(z)+\psi\left(\frac{2}{3}
\right)+\gamma\bigg],
\label{eq_64}
\end{eqnarray}
where $\psi(z)\!=\!\frac{1}{\Gamma(z)}\frac{\mathrm{d}\Gamma(z)}{\mathrm{d}z}$ stands for
a digamma function and $\gamma\!\approx\!0.5772$ is known as the Euler-Mascheroni constant.
The first derivative of the Gaussian hypergeometric function performed with respect to the
parameter $a$, which appears on the right-hand side of Eq.~(\ref{eq_64}), can be determined
according to the following formula
\begin{equation}
\frac{\partial}{\partial a}\pFq{2}{1}\left(a,b,a+1,z\right)=\frac{bz}{(a+1)^{2}}\,
\pFq{3}{2}\left(a+1,a+1,b+1;a+2,a+2;z\right),
\label{eq_65}
\end{equation}
where $\pFq{3}{2}\left(a_{1},a_{2},a_{3};b_{1},b_{2};z\right)$ is a generalized
hypergeometric function. Thus, by applying the last two properties to Eq.~(\ref{eq_63}), we
obtain the final expression for the MFPT downward the harmonic potential in the case of the
non-linear diffusion with the fixed parameter $\sigma\!=\!1$:
\begin{eqnarray}
\mathcal{T}(x_{0},\alpha)&=&\frac{1}{2\alpha}\,\pFq{2}{1}\left(\frac{2}{3},\frac{2}{3},
\frac{5}{3},-\frac{3D}{2\alpha\lvert x_{0}\rvert^{3}}\right)+\frac{D}{3\alpha^{2}\lvert
x_{0}\rvert^{3}}\,\pFq{3}{2}\left(1,1,\frac{5}{3};2,2;-\frac{3D}{2\alpha\lvert x_{0}
\rvert^{3}}\right)
\nonumber\\
&-&\frac{1}{3\alpha}\left[\log\left(\frac{3D}{2\alpha\lvert x_{0}\rvert^{3}}\right)+
\psi\left(\frac{2}{3}\right)+\gamma\right].
\label{eq_66}
\end{eqnarray}

\begin{figure}[t]
\centering
\includegraphics[scale=0.35]{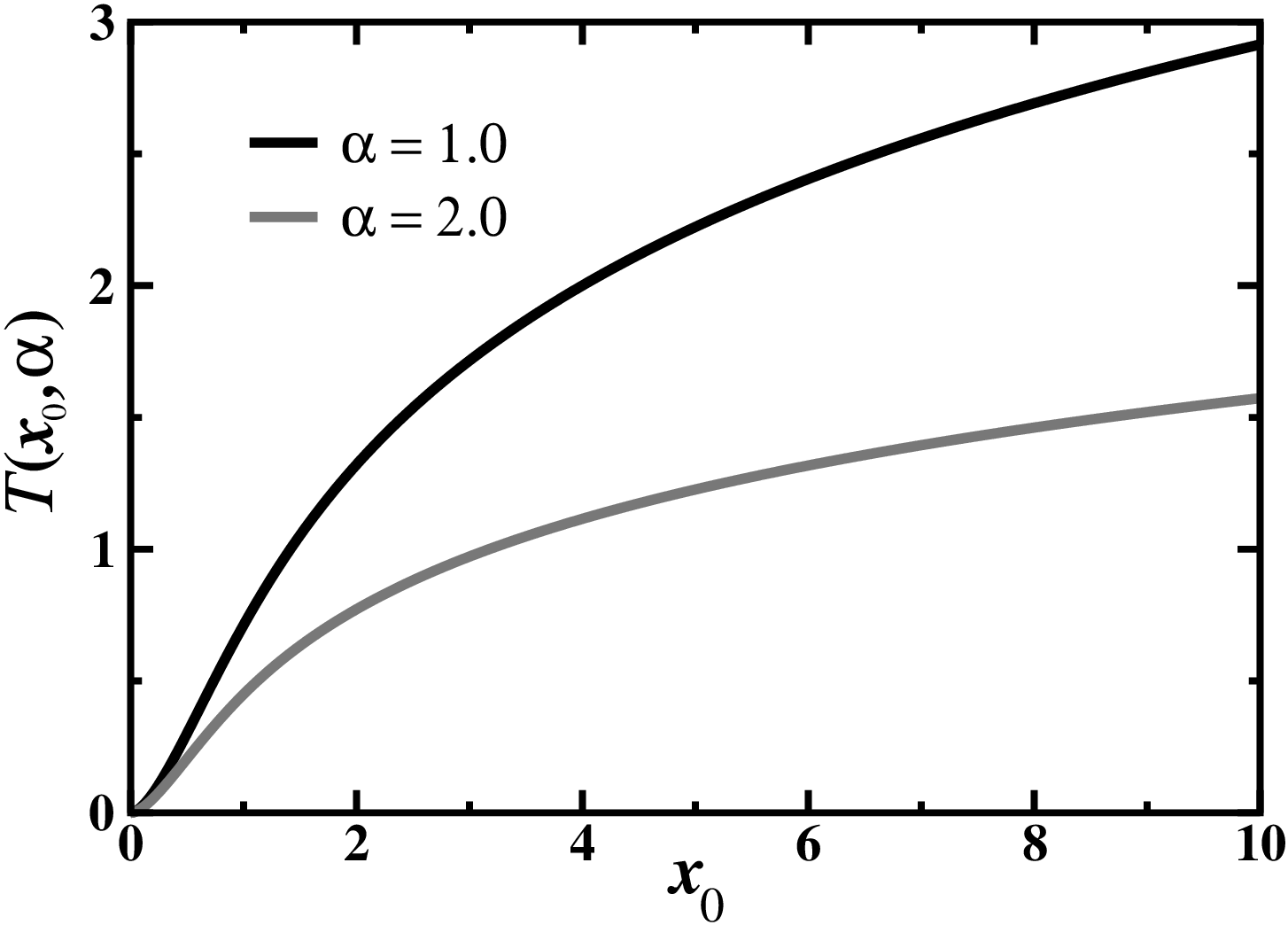}
\caption{Shown is the dependence of the MFPT on the location of the initial position $x_{0}$
relative to the target point placed in the minimum $x\!=\!0$ of the harmonic potential
$V(x)\!=\!\frac{1}{2}\alpha x^{2}$ for the non-linear diffusion given by Eq.~(\ref{eq_54}).
Here, two values of the stiffness $\alpha$ have been chosen and the value of the diffusion
coefficient $D\!=\!1.0$ has been assumed.}
\label{fig13}
\end{figure}
\begin{figure}[t]
\centering
\includegraphics[scale=0.35]{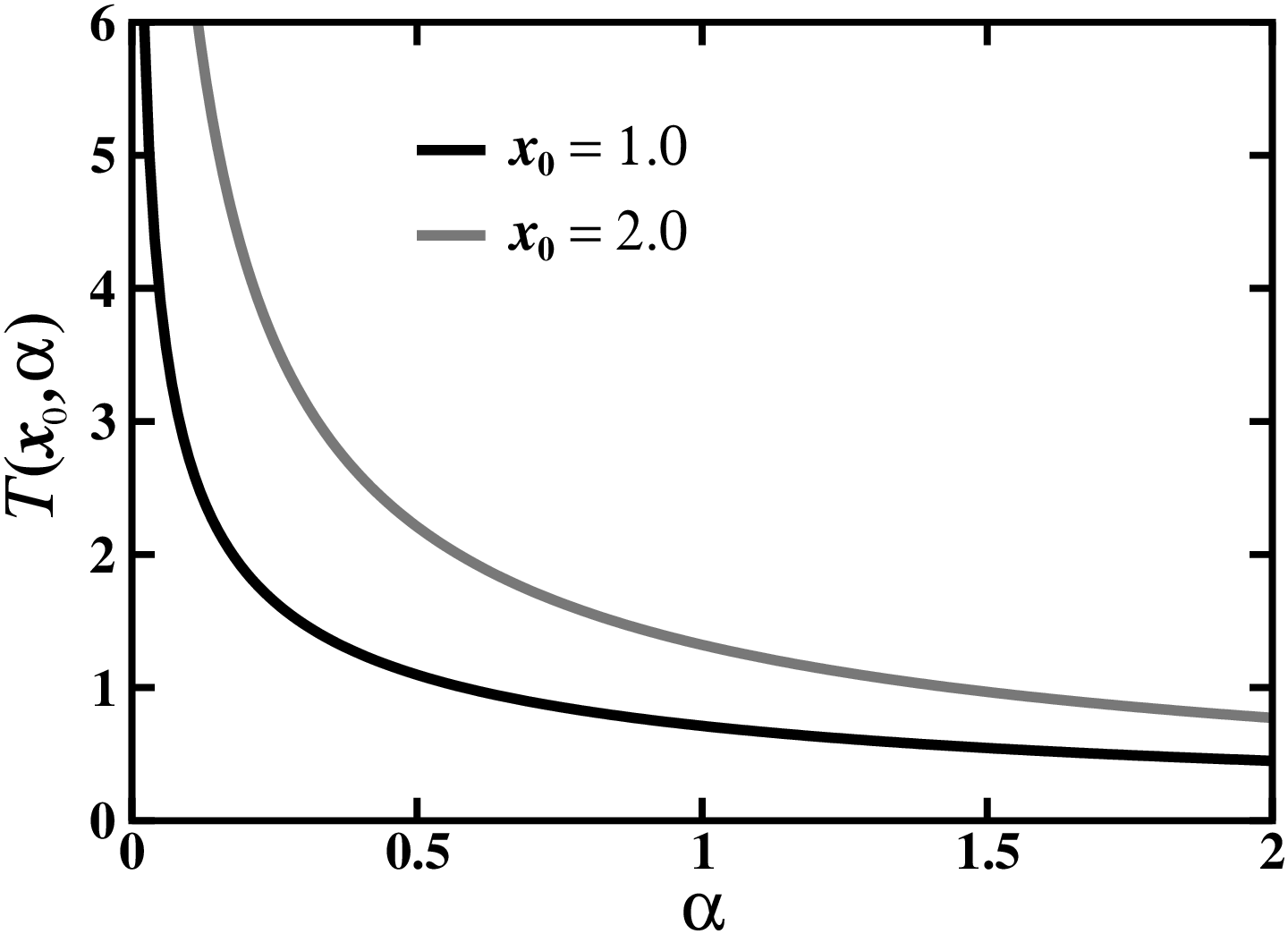}
\caption{Shown is the dependence of the MFPT on the stiffness $\alpha$ of the harmonic potential
$V(x)\!=\!\frac{1}{2}\alpha x^{2}$ for the non-linear diffusion given by Eq.~(\ref{eq_54}).
Here, two values of the initial position $x_{0}$ relative to the target point placed in the
minimum $x\!=\!0$ of this potential have been chosen and the value of the diffusion coefficient
$D\!=\!1.0$ has been assumed.}
\label{fig14}
\end{figure}

\noindent This is the central result of the present section. Fig.~\ref{fig13} illustrates how
the MFPT to the target at $x\!=\!0$ of the harmonic potential depends on the stiffness parameter
$\alpha$. We see that the larger value of $\alpha$ the shorter MFPT for a given distance from
the initial position $x_{0}$ to the target point. In Fig.~\ref{fig14} the dependence of the MFPT
on the stiffness parameter is displayed for two fixed initial positions. We see that for
$\alpha\!\to\!0$ and independently of $x_{0}$ the MFPT tends to infinity, as it should be in
the general case of any diffusive process, and in particular the non-linear diffusion occurred
in the unbounded space. In turn, if the initial position $x_{0}\!\to\!0$ then the MFPT should
vanish. To examine this effect we need to demonstrate that $\lim_{x_{0}\to0}\mathcal{T}
(x_{0},\alpha)\!=\!0$ in Eq.~(\ref{eq_66}). For this purpose, it is enough to prove that the
following limit
\begin{equation}
\lim_{x_{0}\to0}\left[\frac{D}{3\alpha^{2}\lvert x_{0}\rvert^{3}}\,
\pFq{3}{2}\left(1,1,\frac{5}{3};2,2;-\frac{3D}{2\alpha\lvert x_{0}\rvert^{3}}\right)-
\frac{1}{3\alpha}\log\left(\frac{3D}{2\alpha\lvert x_{0}\rvert^{3}}\right)\right]=
\frac{1}{3\alpha}\left[\psi\left(\frac{2}{3}\right)+\gamma\right],
\label{eq_67}
\end{equation}
is correct and then exploit the fact that
\begin{equation}
\lim_{x_{0}\to0}\pFq{2}{1}\left(\frac{2}{3},\frac{2}{3},\frac{5}{3},-\frac{3D}
{2\alpha\lvert x_{0}\rvert^{3}}\right)=0.
\label{eq_68}
\end{equation}
To prove the relation in Eq.~(\ref{eq_67}), we take advantage of the asymptotic representation
of the generalized hypergeometric function for $\lvert z\rvert\!\to\!\infty$, i.e. when
$x_{0}\!\to\!0$ for fixed $D$ and $\alpha$:
\begin{eqnarray}
\pFq{3}{2}\left(a_{1},a_{1},a_{3};b_{1},b_{2};-z\right)&\propto&\frac{\Gamma(b_{1})
\Gamma(b_{2})\Gamma(a_{3}-a_{1})}{\Gamma(a_{1})\Gamma(a_{3})\Gamma(b_{1}-a_{1})
\Gamma(b_{2}-a_{1})z^{a_{1}}}[\log(z)+\psi(a_{3}-a_{1})\\
\nonumber
&-&\psi(b_{1}-a_{1})-\psi(b_{2}-a_{1})-\psi(a_{1})-2\gamma]
+\mathcal{O}\left(\frac{1}{z}\right).
\label{eq_69}
\end{eqnarray}
For the special values of the parameters $a_{1}\!=\!1$, $a_{3}\!=\!\frac{5}{3}$ and
$b_{1}\!=\!b_{2}\!=\!2$, we obtain from above that
\begin{equation}
\pFq{3}{2}\left(1,1,\frac{5}{3};2,2;-\frac{3D}{2\alpha\lvert x_{0}\rvert^{3}}\right)\propto
\frac{\alpha\lvert x_{0}\rvert^{3}}{D}\left[\log\left(\frac{3D}{2\alpha\lvert x_{0}\rvert^{3}}
\right)+\psi\left(\frac{2}{3}\right)+\gamma\right].
\label{eq_70}
\end{equation}
This result implies that the relationship in Eq.~(\ref{eq_67}) is really satisfied. Thus,
we can apply it along with Eq.~(\ref{eq_68}) to Eq.~(\ref{eq_66}) and show that the MFPR for
the non-linear diffusion equation of the harmonically trapped particle disappears when the
initial position $x_{0}$ coincides with the target point in the minimum $x\!=\!0$ of the
harmonic potential.

\section{\label{sec_6}Conclusions}

It is probably no exaggeration to say that in most of the papers devoted to the studies of
first-passage properties of diffusive motion one usually assumes that this process is modeled
in terms of the {\it linear} partial differential equations. Inspired by such a state of
matters, we have centred our efforts to analyse in the present paper the fundamental aspects
of first-passage statistics for the {\it non-linear} diffusion equation. To be more concrete,
we have considered here the special variant of this equation, known as the {\it porous medium
equation}, in which the diffusion coefficient is power-law dependent on the probability density
function of a diffusing particle. Additionally, the corresponding power-law exponent was served
as the non-negative and constant parameter. As stated in Introduction, the porous medium equation
has found useful applications in such different fields as plasma physics, geophysics and biology.

At the beginning, we briefly described the basic properties of the non-linear diffusion
equation along with its typical solution having the form of the Zel'dovitch-Barenblatt-Pattle
algebraic function. Further, the key concepts of the first-passage formalism such as the
survival probability, the first-passage time distribution and the mean first-passage time have
been concisely revised. In the subsequent two sections we determined these quantities for
the ordinary as well as non-linear diffusion that occurred in the semi-infinite interval with
the target point located at the origin. While the use of the image method raises no objections
in the first case, this technique cannot be applied in the latter case due to the non-linear
form of the diffusion equation. Therefore, we had resort to the standard procedure according
to which the probability density function undergoes the absorbing boundary conditions at the
target point, where it disappears. Although we found the exact formula for the survival
probability, it has appeared that this quantity does not meet the crucial property, namely,
the initial condition. Then, we have argued that the solution of the non-linear diffusion
equation in the presence of the absorbing well entails disappearance of the probability
current. Consequently, the totally absorbing well acts as a perfectly reflecting wall and
this contradiction leads to ambiguous situation. Therefore, we have chosen an alternative
method where instead of the absorbing target there appears the target point to which a diffusing
particle arrives for the first time and then can return to it many times.

In this way, we were able to obtain the exact and approximate results for the survival
probability and hence the first-passage time distribution. The former concern the power-law
exponent $\sigma\!=\!1$ and $2$, whereas the latter correspond to its values in the range
between $1$ and $2$. Moreover, the approximate formulae, even though described by completely
different expressions, perfectly agree with the exact results. We have also shown that the
time course of the survival probability for the free non-linear diffusion takes place in two
consecutive stages. For the initial period of time its value remains equal to unity and only
in the second phase of diffusive motion permanently decreases to reach zero in the
infinite-time limit. In turn, the first-passage time distribution is always normalized to
unity although its first moment, that is the mean first-passage time to the target, diverges
to infinity. Such a tendency changes when the non-linear diffusion occurs in the confining
potential. We have shown this on the example of the harmonically trapped particle that
diffuses downward the potential to reach the target located in the minimum. The first-passage
properties of the non-linear diffusion in external potentials will be continued in the future
paper.

The exploration of first-passage phenomena still attracts unabated attention among scientific
community. Despite the immense literature on this subject, our understanding of first-passage
dynamics remains incomplete and requires further systematic and in-depth studies. We hope that
the present paper has become an inherent part of these investigations, specifically, in the
scope of diffusive processes.


\section*{Conflicts of Interest}
The author declares no conflict of interest.
\section*{ORCID iD}
Przemys\l{}aw Che\l{}miniak, https://orcid.org/0000-0002-0085-9232
\appendix
\section{The Laplace transform of PDF for the parameter $\boldsymbol{\sigma\!=\!2}$}

The purpose of this supplementary section is to present the detailed derivation of
Eq.~(\ref{eq_37}) comprised in the main text. Setting the parameter $\sigma\!=\!2$, we get
from Eqs.~(\ref{eq_7}) and (\ref{eq_8}) that two numerical factors appearing in Eq.~(\ref{eq_35})
take the following values, $\mathcal{A}(2)\!=\!\frac{1}{\sqrt{\pi}}$ and $\mathcal{B}(2)\!=\!
\frac{\pi}{4}$. In addition, the auxiliary function in this equation is $\xi(x,2)\!=\!\left(
\frac{\pi}{4}\right)^{2}\frac{(x-x')^{4}}{D}$. Thus, inserting all these quantities into 
Eq.~(\ref{eq_35}), we find that the Laplace transform of the PDF, or the propagator
$p(x,t\!\mid\!x')$ from $x'$ to $x$ at time $t$, is
\begin{equation}
\tilde{p}(x,s\!\mid\!x')=\frac{1}{\sqrt{\pi}}\bigintsss_{\xi(x,2)}^{\infty}
\frac{\mathrm{e}^{-st}}{(Dt)^{\frac{1}{4}}}\sqrt{1-\frac{\pi(x-x')^{2}}
{4\sqrt{Dt}}}\,\mathrm{d}t
\label{eq_A1}
\end{equation}
Now, upon introducing the new notation $a(x)\!\equiv\!\sqrt{\xi(x,2)}$ and changing the
variable of integration from $t$ to $\tau$, so that $t\!=\!(\tau^{2}+a(x))^{2}$, we can recast
the Laplace transform of PDF in Eq.~(\ref{eq_A1}) to the much simpler form:
\begin{equation}
\tilde{p}(x,s\!\mid\!x')=\frac{4}{\sqrt{\pi\sqrt{D}}}\int_{0}^{\infty}\tau^{2}\exp\left[
-s(\tau^{2}+a(x))^{2}\right]\,\mathrm{d}\tau.
\label{eq_A2}
\end{equation}
From now on, the rest of our calculations boils down to find the solution of the above integral.
It can be obtained by using the similarly looking integral 
\begin{equation}
\int_{0}^{\infty}\exp\left(-\alpha\tau^{4}-2\beta\tau^{2}\right)\mathrm{d}\tau=
\frac{1}{2}\sqrt{\frac{\beta}{2\alpha}}\exp\left(\frac{\beta^{2}}{2\alpha}\right)
\mathrm{K}_{\frac{1}{4}}\left(\frac{\beta^{2}}{2\alpha}\right),
\label{eq_A3}
\end{equation}
where the factors $\alpha$, $\beta\!>\!0$, while $\mathrm{K}_{\nu}(z)$ is the modified
Bessel function of the second kind. Out of many well-known properties of this function
we utilize the two ones:
\begin{equation}
\mathrm{K}_{\nu}(z)=\mathrm{K}_{-\nu}(z),
\label{eq_A4}
\end{equation}
and
\begin{equation}
\frac{\mathrm{d}}{\mathrm{d}z}\mathrm{K}_{\nu}(z)=-\mathrm{K}_{\nu-1}(z)-\frac{\nu}{z}
\mathrm{K}_{\nu}(z).
\label{eq_A5}
\end{equation}
First, however, let us differentiate the left and the right side of Eq.~(\ref{eq_A3}) with
respect to the parameter $\beta$. In this way, we have
\begin{equation}
\int_{0}^{\infty}\tau^{2}\exp\left(-\alpha\tau^{4}-2\beta\tau^{2}\right)\mathrm{d}\tau=
-\frac{1}{4\sqrt{2\alpha}}\frac{\mathrm{d}}{\mathrm{d}\beta}\left[\sqrt{\beta}
\exp\left(\frac{\beta^{2}}{2\alpha}\right)\mathrm{K}_{\frac{1}{4}}\left(\frac{\beta^{2}}
{2\alpha}\right)\right].
\label{eq_A6}
\end{equation}
Taking into account Eqs.~(\ref{eq_A4}) and (\ref{eq_A5}), we show that a derivative of
the expression enclosed in the square bracket on the right hand side of the above equation
\begin{equation}
\frac{\mathrm{d}}{\mathrm{d}\beta}\left[\sqrt{\beta}
\exp\left(\frac{\beta^{2}}{2\alpha}\right)\mathrm{K}_{\frac{1}{4}}\left(\frac{\beta^{2}}
{2\alpha}\right)\right]=\frac{\beta\sqrt{\beta}}{\alpha}\exp\left(\frac{\beta^{2}}{2\alpha}
\right)\left[\mathrm{K}_{\frac{1}{4}}\left(\frac{\beta^{2}}{2\alpha}\right)-
\mathrm{K}_{\frac{3}{4}}\left(\frac{\beta^{2}}{2\alpha}\right)\right].
\label{eq_A7}
\end{equation}
If we plugin this derivative back into Eq.~(\ref{eq_A6}) and multiply their both sides by
the exponential function $\exp(-\frac{\beta^{2}}{\alpha})$, we obtain that the integral
\begin{equation}
\bigintsss_{0}^{\infty}\tau^{2}\exp\left[-\alpha\left(\tau^{2}+\frac{\beta}{\alpha}\right)^{2}
\right]\mathrm{d}\tau=\frac{1}{4\sqrt{2}}\left(\frac{\beta}{\alpha}\right)^{\frac{3}{2}}
\exp\left(-\frac{\beta^{2}}{2\alpha}\right)\left[\mathrm{K}_{\frac{3}{4}}\left(
\frac{\beta^{2}}{2\alpha}\right)-\mathrm{K}_{\frac{1}{4}}\left(\frac{\beta^{2}}{2\alpha}
\right)\right].
\label{eq_A8}
\end{equation}
Having this equation at our disposal and assuming that $\alpha\!=\!s$ and $\beta\!=\!sa(x)$,
we immediately solve the integral in Eq.~(\ref{eq_A2}). The final result is as follows
\begin{equation}
\tilde{p}(x,s\!\mid\!x')=\frac{a^{\frac{3}{2}}(x)}{\sqrt{2\pi\sqrt{D}}}\exp\!
\left(-\frac{sa^{2}(x)}{2}\right)\left[\mathrm{K}_{\frac{3}{4}}\!\left(\frac{sa^{2}(x)}{2}
\right)-\mathrm{K}_{\frac{1}{4}}\!\left(\frac{sa^{2}(x)}{2}\right)\right],
\label{eq_A9}
\end{equation}
where the variable $a(x)\!=\!\frac{\pi(x-x')^{2}}{4\sqrt{D}}$.

To complete this supplement, we should also determine the Laplace transform of PDF for
$x\!=\!x'$. It can be done in two ways. The first method takes advantage of the following
limit values of the modified Bessel functions of the second kind:
\begin{equation}
\lim_{z\to0}z^{3}\mathrm{K}_{\frac{1}{4}}\left(uz^{4}\right)=0,
\label{eq_A10}
\end{equation}
and
\begin{equation}
\lim_{z\to0}z^{3}\mathrm{K}_{\frac{3}{4}}\left(uz^{4}\right)=
\frac{\Gamma\left(\frac{3}{4}\right)}{\sqrt[4]{2}\,u^{3/4}}.
\label{eq_A11}
\end{equation}
Calculating these limits in Eq.~(\ref{eq_A9}) we easily show that
\begin{equation}
\tilde{p}(0,s\!\mid\!0)=\frac{\Gamma\left(\frac{3}{4}\right)}{\sqrt{\pi\sqrt{D}}\,
s^{\frac{3}{4}}}.
\label{eq_A12}
\end{equation}
The second method that guarantees the above result consists in the direct calculation of the integral in Eq.~(\ref{eq_A2}) for $x\!=\!x'$ with $\xi(x',2)\!=\!0$. In this case
we use the integral representation of the Euler gamma function, i.e.
$\Gamma(\alpha)\!=\!\int_{0}^{\infty}u^{\alpha-1}\mathrm{e}^{-u}\mathrm{d}u$.

\section{The inverse Laplace transformation of the product of power-law and incomplete
gamma functions}

In this supplementary section we show how to derive the inverse Laplace transform of the
function being the composition of the power-law and the upper incomplete gamma functions.
The Laplace transforms of these functions are well known individually and can be found
in Ref.~\cite{Ober1973}.

To begin with, let us consider the integral representation of the Gaussian hypergeometric
function~\cite{Slat1966}
\begin{equation}
\pFq{2}{1}(a,b,c,z)=\frac{\Gamma(c)}{\Gamma(b)\Gamma(c-b)}\int_{0}^{1}\nu^{b-1}
(1-\nu)^{c-b-1}(1-z\nu)^{-a}\mathrm{d}\nu,
\label{eq_B1}
\end{equation}
where $a$, $b$ and $c$ are real numbers and $\Gamma(z)$ is the Euler gamma function. Upon
making the following substitution, where $a\!=\!1$, $b\!=\!1\!-\!\beta$,
$c\!=\!2\!-\!(\alpha\!+\!\beta)$ and $z\!=\!1\!-\!\frac{1}{u}$, we have
\begin{equation}
\pFq{2}{1}\left(1,1-\beta,2-(\alpha+\beta),1-u^{-1}\right)=\frac{1}{\mathrm{B}
(1-\alpha,1-\beta)}\int_{0}^{1}\frac{u\,\mathrm{d}\nu}{\nu^{\beta}(1-\nu)^{\alpha}
[\nu+u(1-\nu)]},
\label{eq_B2}
\end{equation}
where $\mathrm{B}(x,y)\!=\!\frac{\Gamma(x)\Gamma(y)}{\Gamma(x+y)}$ is the Euler beta function.
By defining the new variable of integration $\xi\!=\!\nu+u(1-\nu)$, we can easily convert the
right hand side of Eq.~(\ref{eq_B2}) into the following form:
\begin{equation}
\pFq{2}{1}\left(1,1-\beta,2-(\alpha+\beta),1-u^{-1}\right)=\frac{u(1-u)^{\alpha+\beta-1}}
{\mathrm{B}(1-\alpha,1-\beta)}\int_{u}^{1}\frac{\mathrm{d}\xi}{\xi(1-\xi)^{\alpha}
(\xi-u)^{\beta}}.
\label{eq_B3}
\end{equation}
The above expression determines the useful integral
\begin{equation}
\int_{u}^{1}\frac{\mathrm{d}\xi}{\xi(1-\xi)^{\alpha}(\xi-u)^{\beta}}=
\frac{\mathrm{B}(1-\alpha,1-\beta)\,\pFq{2}{1}\left(1,1-\beta,2-(\alpha+\beta),1-u^{-1}
\right)}{u(1-u)^{\alpha+\beta-1}},
\label{eq_B4}
\end{equation}
which will turn out to be essential in the context of our further argumentation.

To complete the task, we will proceed as follows. Let $\mathcal{L}[f(t);t]$ denotes the
Laplace transformation of a function $f(t)$ and let $\mathcal{L}^{-1}[\tilde{f}(s);s]$ stands
for the inverse transformation of the Laplace transform
$\tilde{f}(s)\!:=\int_{0}^{\infty}f(t)\mathrm{e}^{-st}\mathrm{d}t$. According to Ref.
~\cite{Ober1973}, the inverse transformation of the Laplece transform
$\tilde{f}(s)\!=\!\Gamma(\beta,a\,s)$ of the upper incomplete gamma function reads
\begin{equation}
\mathcal{L}^{-1}[\Gamma(\beta,a\,s);s]=\frac{a^{\beta}\,\Theta(t-a)}{\Gamma(1-\beta)
t(t-a)^{\beta}},
\label{eq_B5}
\end{equation}
where $a\!>\!0$ and $\mathrm{Re}(\beta)\!<\!1$. Here, $\Theta(z)$ is the Heviside unit step
function equal to 1, if $z\geqslant0$ and 0 otherwise. In turn, if $f(t)\!=\!t^{-\alpha}$ for
$\alpha\!<\!1$ then $\tilde{f}(s)\!=\!\Gamma(1\!-\!\alpha)\,s^{\alpha-1}$. By virtue of the
convolution theorem, which states that the Laplace transformation of the convolution
$f(t)\!\ast\!g(t)\!:=\!\int_{0}^{t}f(\tau)g(t-\tau)\mathrm{d}\tau$ of two integrable functions
$f(t)$ and $g(t)$ is the product of their Laplace transforms, i.e.
$\mathcal{L}[f(t)\!\ast\!g(t);t]\!=\!\tilde{f}(s)\tilde{g}(s)$, we can express the inverse
Laplace transformation of the power-law
function $\tilde{f}(s)\!=\!s^{\alpha-1}$ and the upper incomplete gamma function
$\tilde{g}(s)\!=\!\Gamma(\beta,as)$, both defined in the Laplace domain, as follows:
\begin{align}
\mathcal{L}^{-1}[s^{\alpha-1}\Gamma(\beta,a\,s);s]&=\int_{0}^{t}\frac{a^{\beta}\,
\Theta(\tau-a)}{\Gamma(1-\alpha)\Gamma(1-\beta)\tau(\tau-a)^{\beta}(t-\tau)^{\alpha}}\,
\mathrm{d}\tau\notag\\
&=\frac{a^{\beta}}{\Gamma(1-\alpha)\Gamma(1-\beta)}\int_{a}^{t}\frac{\mathrm{d}\tau}{
\tau(t-\tau)^{\alpha}(\tau-a)^{\beta}}.
\label{eq_B6}
\end{align}
Assuming the new variable of integration $\zeta\!=\!\frac{\tau}{t}$ and denoting
$u\!=\!\frac{a}{t}$, we recast the integral and the remaining part of the formula in the
second line of Eq.~(\ref{eq_B6}) as
\begin{equation}
\mathcal{L}^{-1}[s^{\alpha-1}\Gamma(\beta,a\,s);s]=\frac{a^{\beta}t^{-(\alpha+\beta)}}{
\Gamma(1-\alpha)\Gamma(1-\beta)}\int_{u}^{1}\frac{\mathrm{d}\zeta}{\zeta(1-\zeta)^{\alpha}
(\zeta-u)^{\beta}}.
\label{eq_B7}
\end{equation}
Let us note that the integral appearing in the above equation has exactly the same form as
the integral determined in Eq.~(\ref{eq_B4}). Hence, we conclude that the inverse Laplace
transformation of the product of the power-law and the upper incomplete gamma functions is
given by
\begin{equation}
\mathcal{L}^{-1}[s^{\alpha-1}\Gamma(\beta,a\,s);s]=\left(\frac{a}{t}\right)^{\beta-1}
\left(1-\frac{a}{t}\right)^{1-(\alpha+\beta)}\frac{\pFq{2}{1}\left(1,1-\beta,
2-(\alpha+\beta),1-\frac{t}{a}\right)}{t^{\alpha}\,\Gamma(2-(\alpha+\beta))}.
\label{eq_B8}
\end{equation}
We can use the general result in Eq.~(\ref{eq_B8}) to determine the survival probability for
the particular value of the parameter $\sigma\!=\!1$ (see the main text). For this purpose,
it is enough to set $\alpha\!=\!0$ and $\beta\!=\!\frac{2}{3}$, which gives from Eq.~(\ref{eq_B8}) that
\begin{equation}
\mathcal{L}^{-1}\left[\frac{1}{s}\,\Gamma\left(\frac{2}{3},a\,s\right);s\right]=
\left(\frac{t}{a}-1\right)^{\frac{1}{3}}\frac{\pFq{2}{1}\left(1,\frac{1}{3},\frac{4}{3},
1-\frac{t}{a}\right)}{\Gamma\left(\frac{4}{3}\right)}.
\label{eq_B9}
\end{equation}
On the other hand, we have to assume $\alpha\!=\!\frac{2}{3}$ and $\beta\!=\!0$, which gives that
\begin{equation}
\mathcal{L}^{-1}\left[s^{-\frac{1}{3}}\Gamma\left(0,a\,s\right);s\right]=
\left(\frac{t}{a}-1\right)^{\frac{1}{3}}\frac{\pFq{2}{1}\left(1,1,\frac{4}{3},1-\frac{t}{a}
\right)}{a^{\frac{2}{3}}\,\Gamma\left(\frac{4}{3}\right)}.
\label{eq_B10}
\end{equation}
The survival probability is by definition a real and non-negative quantity. To ensure the
fulfilment of this condition, we have to additionally require that $t\geqslant a$. Therefore,
it is enough to complete all the three last formulae for the Laplace inverse transformation
by multiplying their right hand sides by the Heviside unit step function $\Theta(t\!-\!a)$.

\section{The inverse Laplace transformation of the product of power-law, exponential and
modified Bessel functions}

Suppose again that $\mathcal{L}[f(t);t]$ denotes the Laplace transformation of a function
$f(t)$ and let $\mathcal{L}^{-1}[\tilde{f}(s);s]$ stands for the inverse transformation of the
Laplace transform $\tilde{f}(s)\!:=\int_{0}^{\infty}f(t)\mathrm{e}^{-st}\mathrm{d}t$. The
inverse Laplace transformation of $\tilde{f}(s)$, when additionally multiplied by the
exponential function $\mathrm{e}^{-bs}$, defined in the Laplace domain with $b\!>\!0$, is
\begin{equation}
\mathcal{L}^{-1}\left[\mathrm{e}^{-bs}f(s);s\right]=\Theta(t-b)f(t-b),
\label{eq_C1}
\end{equation}
where, as before, $\Theta(z)$ means the Heviside unit step function equal to 1, if $z\geqslant0$
and 0 otherwise. According to Ref.~\cite{Ober1973} the inverse Laplace transformation of the
power-law function $s^{-\nu}$ multiplied by the modified Bessel function of the second kind
$\mathrm{K}_{\nu}(bs)$ with $b\!>\!0$ is as follows:
\begin{equation}
\mathcal{L}^{-1}\left[s^{-\nu}\mathrm{K}_{\nu}(bs);s\right]=\Theta(t-b)\sqrt{\frac{\pi}{2b}}
\left(t^{2}-b^{2}\right)^{\frac{\mu}{2}-\frac{1}{4}}P_{\nu-\frac{1}{2}}^{\frac{1}{2}-\mu}
\left(\frac{t}{b}\right),
\label{eq_C2}
\end{equation}
where $P_{\beta}^{\alpha}(z)$ corresponds to the associated Legendre function~\cite{Grad2007}.
In the case when $z\!>\!1$, this special function can be represented by the Gaussian
hypergeometric function according to the following formula:
\begin{equation}
P_{\beta}^{\alpha}(z)=\frac{1}{\Gamma(1-\alpha)}\left(\frac{z+1}{z-1}\right)
^{\frac{\alpha}{2}}\pFq{2}{1}\left(-\beta,\beta+1,1-\alpha,\frac{1-z}{2}\right).
\label{eq_C3}
\end{equation}
Now, combining Eqs.~(\ref{eq_C1}) and (\ref{eq_C2}) and conducting elementary calculations,
we obtain that
\begin{equation}
\mathcal{L}^{-1}\left[s^{-\nu}\mathrm{K}_{\nu}(bs)\mathrm{e}^{-bs};s\right]=\Theta(t-2b)
\sqrt{\frac{\pi}{2b}}\left[(t-b)^{2}-b^{2}\right]^{\frac{\mu}{2}-\frac{1}{4}}
P_{\nu-\frac{1}{2}}^{\frac{1}{2}-\mu}\left(\frac{t}{b}-1\right).
\label{eq_C4}
\end{equation}
Next, it is enough to insert Eq.~(\ref{eq_C3}) into the above formula to get the final result:
\begin{equation}
\mathcal{L}^{-1}\left[s^{-\nu}\mathrm{K}_{\nu}(bs)\mathrm{e}^{-bs};s\right]=\Theta(t-2b)
\sqrt{\frac{\pi}{2b}}\frac{(t-2b)^{\mu-\frac{1}{2}}}{\Gamma\left(\mu+\frac{1}{2}\right)}
\pFq{2}{1}\left(\frac{1}{2}-\nu,\frac{1}{2}+\nu,\frac{1}{2}+\mu,1-\frac{t}{2b}\right).
\label{eq_C5}
\end{equation}
To determine the inverse Laplace transformation of the survival probability in Eq.~(\ref{eq_39}),
we have to set $\mu\!=\!\frac{1}{4}$ and, respectively, $\nu\!=\!\frac{1}{4}$ and
$\nu\!=\!\frac{3}{4}$. In the first case we obtain from Eq.~(\ref{eq_C5}) that
\begin{eqnarray}
\mathcal{L}^{-1}\left[s^{-\frac{1}{4}}\mathrm{K}_{\frac{1}{4}}(a^{2}s)\mathrm{e}^{-a^{2}s};s
\right]&=&\Theta(t-2a^{2})\sqrt{\frac{\pi}{2}}\,\frac{\pFq{2}{1}\left(\frac{1}{4},\frac{3}{4},
\frac{3}{4},1\!-\!\frac{t}{2a^{2}}\right)}{a\,\Gamma\left(\frac{3}{4}\right)\left(t-2a^{2}
\right)^{\frac{1}{4}}}\nonumber\\
&=&\Theta(t-2a^{2})\sqrt{\frac{\pi}{a\sqrt{2}}}\,\frac{1}{\Gamma\left(\frac{3}{4}\right)
[t\left(t-2a^{2}\right)]^{\frac{1}{4}}},
\label{eq_C6}
\end{eqnarray}
where the formula in the second line results from the fact that $\pFq{2}{1}\left(\frac{1}{4},
\frac{3}{4},\frac{3}{4},z\right)\!=\!(1-z)^{-1/4}$, whereas in the second case
\begin{equation}
\mathcal{L}^{-1}\left[s^{-\frac{1}{4}}\mathrm{K}_{\frac{3}{4}}(a^{2}s)\mathrm{e}^{-a^{2}s};s
\right]=\Theta(t-2a^{2})\sqrt{\frac{\pi}{2}}\,\frac{\pFq{2}{1}\left(-\frac{1}{4},\frac{5}{4},
\frac{3}{4},1\!-\!\frac{t}{2a^{2}}\right)}{a\,\Gamma\left(\frac{3}{4}\right)\left(t-2a^{2}
\right)^{\frac{1}{4}}}.
\label{eq_C7}
\end{equation}
In both the above expressions, the variable $a\!=\!\frac{\pi x^{2}}{4\sqrt{2D}}$ depends on the
position of the diffusing particle.

\section{Exact time derivatives of survival probabilities}

The objective of the present addendum is to figure out the first derivatives of survival
probabilities with respect to time provided the parameter $\sigma\!=\!1$ and $2$. These
results correspond in fact to the exact expressions for the first-passage time distributions
contained in Eqs.~(\ref{eq_42}) and (\ref{eq_43}) of the main text.

Let us first consider the case of the parameter $\sigma\!=\!1$, for which the survival
probability is defined in Eq.~(\ref{eq_40}). Conducting precise calculus we find that its
first derivative over the time preceded by the negative sign is as follows:
\begin{eqnarray}
-\frac{\mathrm{d}Q_{1}(t\!\mid\!x_{0})}{\mathrm{d}t}&=&\frac{\sqrt{3}}{2\pi t}\,
\Theta\left(t-\tau_{1}(x_{0})\right)\left(\frac{t}{\tau_{1}(x_{0})}-1\right)^{-\frac{2}{3}}
\frac{t}{\tau_{1}(x_{0})}
\bigg[\pFq{2}{1}\left(1,\frac{1}{3},\frac{4}{3},1-\frac{t}{\tau_{1}(x_{0})}\right)\nonumber\\
&-&\pFq{2}{1}\left(1,1,\frac{4}{3},1-\frac{t}{\tau_{1}(x_{0})}\right)+\frac{9}{4}
\left(\frac{t}{\tau_{1}(x_{0})}-1\right)\pFq{2}{1}\left(2,2,\frac{7}{3},1-\frac{t}
{\tau_{1}(x_{0})}\right)\nonumber\\
&-&\frac{3}{4}\left(\frac{t}{\tau_{1}(x_{0})}-1\right)\pFq{2}{1}\left(2,\frac{4}{3},
\frac{7}{3},1-\frac{t}{\tau_{1}(x_{0})}\right)\bigg],
\label{eq_D1}
\end{eqnarray}
where the auxiliary function $\tau_{1}(x_{0})\!=\!\frac{2\lvert x_{0}\rvert^{3}}{9Dt}$.
We show farther how to simplify this highly complex formula. For this purpose, we use the
functional identities which reflect intrinsic properties of the Gaussian hypergeometric functions. The first two identities constitute the system of the following equations:
\begin{equation}
c\,\pFq{2}{1}(a-1,b,c,z)+c\,(z-1)\,\pFq{2}{1}(a,b,c,z)+(b-c)\,z\,\pFq{2}{1}(a,b,c+1,z)=0,
\label{eq_D2}
\end{equation}
\begin{equation}
(b-c)\,\pFq{2}{1}(a,b-1,c,z)+(c-a-b)\,\pFq{2}{1}(a,b,c,z)=a\,(z-1)\,\pFq{2}{1}(a+1,b,c,z).
\label{eq_D3}
\end{equation}
By making the change $c\!\to\!c-1$ in Eq.~(\ref{eq_D2}) and $a\!\to\!a-1$ in Eq.~(\ref{eq_D3}),
we can combine these two relations to have the single identity
\begin{equation}
\pFq{2}{1}(a-1,b-1,c-1,z)-\pFq{2}{1}(a-1,b,c-1,z)=-\frac{a-1}{c-1}\,z\,\pFq{2}{1}(a,b,c,z).
\label{eq_D4}
\end{equation}
Setting the particular values for $a\!=\!2$, $b\!=\!2$ and $c\!=\!7/3$, we obtain from
the above formula that
\begin{equation}
\pFq{2}{1}\left(1,1,\frac{4}{3},z\right)-\pFq{2}{1}\left(1,2,\frac{4}{3},z\right)
=-\frac{3}{4}\,z\,\pFq{2}{1}\left(2,2,\frac{7}{3},z\right).
\label{eq_D5}
\end{equation}
The second system of equations is established by the next two functional identities
\begin{equation}
c\,\pFq{2}{1}(a,b-1,c,z)+c\,(z-1)\,\pFq{2}{1}(a,b,c,z)+(a-c)\,z\,\pFq{2}{1}(a,b,c+1,z)=0,
\label{eq_D6}
\end{equation}
\begin{equation}
(a-c)\,\pFq{2}{1}(a-1,b,c,z)+(c-a-b)\,\pFq{2}{1}(a,b,c,z)=b\,(z-1)\,\pFq{2}{1}(a,b+1,c,z)=0.
\label{eq_D7}
\end{equation}
Again, converting $c\!\to\!c-1$ in Eq.~(\ref{eq_D6}) and $b\!\to\!b-1$ in Eq.~(\ref{eq_D7}),
we connect these two equations to construct the following identity
\begin{equation}
\pFq{2}{1}(a-1,b-1,c-1,z)-\pFq{2}{1}(a,b-1,c-1,z)=-\frac{b-1}{c-1}\,z\,\pFq{2}{1}(a,b,c,z).
\label{eq_D8}
\end{equation}
By choosing the specific values for $a\!=\!2$, $b\!=\!4/3$ and $c\!=\!7/3$, we see that
Eq.~(\ref{eq_D8}) takes the special form
\begin{equation}
\pFq{2}{1}\left(1,\frac{1}{3},\frac{4}{3},z\right)-\pFq{2}{1}\left(2,\frac{1}{3},
\frac{4}{3},z\right)=-\frac{1}{4}\,z\,\pFq{2}{1}\left(2,\frac{4}{3},\frac{7}{3},z\right).
\label{eq_D9}
\end{equation}
To complete the set of relations embodied by Eqs.~(\ref{eq_D5}) and (\ref{eq_D9}), we take
advantage of the following functional identity
\begin{equation}
(a-c)\,\pFq{2}{1}(a-1,b,c,z)+[c-2a+(a-b)z]\,\pFq{2}{1}(a,b,c,z)=a\,(z-1)\,
\pFq{2}{1}(a+1,b,c,z),
\label{eq_D10}
\end{equation}
and the property according to which if $a\!=\!0$, then $\pFq{2}{1}(0,b,c,z)\!=\!1$. Therefore,
setting in Eq.~(\ref{eq_D10}) $a\!=\!1$, $b\!=\!1/3$, $c\!=\!4/3$ and converting $z\!\to\!1-z$
lead to
\begin{equation}
3\,\pFq{2}{1}\left(2,\frac{1}{3},\frac{4}{3},1-z\right)-2\,\pFq{2}{1}\left(1,\frac{1}{3},
\frac{4}{3},1-z\right)=\frac{1}{z}.
\label{eq_D11}
\end{equation}

If we now multiply both sides of Eqs.~(\ref{eq_D5}) and (\ref{eq_D9}) by $3$ and change $z\!\to\!1-z$, as well as apply Eq.~(\ref{eq_D11}), we readily construct the following
identity relation between the Gaussian hypergeometric functions:
\begin{eqnarray}
&&\pFq{2}{1}\left(1,\frac{1}{3},\frac{4}{3},1-z\right)
-\pFq{2}{1}\left(1,1,\frac{4}{3},1-z\right)+\frac{9}{4}(z-1)\,
\pFq{2}{1}\left(2,2,\frac{7}{3},1-z\right)-\frac{3}{4}(z-1)\nonumber\\
&\times&\pFq{2}{1}\left(2,\frac{4}{3},\frac{7}{3},1-z\right)=\frac{1}{z}+
2\,\pFq{2}{1}\left(1,1,\frac{4}{3},1-z\right)-3\,\pFq{2}{1}\left(1,2,\frac{4}{3},1-z\right).
\label{eq_D12}
\end{eqnarray}
A quick look at Eq.~(\ref{eq_D1}) convinces us that the full expression enclosed in the
square brackets of this formula is exactly the same as the left-hand side of the above
identity with $z\!=\!t/\tau_{1}(x_{0})$. In this way, we can reduce Eq.~(\ref{eq_D1}) to
the much simpler form
\begin{eqnarray}
-\frac{\mathrm{d}Q_{1}(t\!\mid\!x_{0})}{\mathrm{d}t}&=&\frac{\sqrt{3}}{2\pi t}\,
\Theta\left(t-\tau(x_{0})\right)\left(\frac{t}{\tau(x_{0})}-1\right)^{-\frac{2}{3}}
\bigg\{1+\frac{t}{\tau(x_{0})}\nonumber\\
&\times&\bigg[2\,\pFq{2}{1}\left(1,1,\frac{4}{3},1-\frac{t}{\tau(x_{0})}
\right)-3\,\pFq{2}{1}\left(1,2,\frac{4}{3},1-\frac{t}{\tau(x_{0})}\right)\bigg]\bigg\}.
\label{eq_D13}
\end{eqnarray}

Now, we turn to the case of the parameter $\sigma\!=\!2$. Exact calculations of the time
derivative of the survival probability in Eq.~(\ref{eq_41}) preceded by the
negative sign yields
\begin{eqnarray}
-\frac{\mathrm{d}Q_{2}(t\!\mid\!x_{0})}{\mathrm{d}t}&=&\frac{\Gamma^{2}\!\left(\frac{1}{4}
\right)\lvert x_{0}\rvert}{4\sqrt{2}\,\pi D^{1/4}}\,\Theta\!\left(t-\tau_{2}(x_{0})\right)
\bigg\{\frac{1}{4(t-\tau_{2}(x_{0}))^{5/4}}\bigg[\left(\frac{t}{\tau_{2}(x_{0})}\right)
^{-\frac{1}{4}} \nonumber\\
&-&\pFq{2}{1}\left(-\frac{1}{4},\frac{5}{4},\frac{3}{4},1-\frac{t}{\tau_{2}(x_{0})}
\right)\bigg]-\frac{1}{(t-\tau_{2}(x_{0}))^{1/4}}\nonumber\\
&\times&\frac{\partial}{\partial t}\bigg[
\left(\frac{t}{\tau_{2}(x_{0})}\right)^{-\frac{1}{4}}
-\pFq{2}{1}\left(-\frac{1}{4},\frac{5}{4},\frac{3}{4},1-\frac{t}{\tau_{2}(x_{0})}
\right)\bigg]\bigg\},
\label{eq_D14}
\end{eqnarray}
where $\tau_{2}(x_{0})\!=\!\frac{\pi^{2}x_{0}^{4}}{16D}$. To determine the time derivative
in the last line of the above formula, we use the system of two equations
\begin{equation}
\frac{\partial}{\partial z}\left[z^{a}\pFq{2}{1}(a,b,c,z)\right]=a\,z^{a-1}
\pFq{2}{1}(a+1,b,c,z)
\label{eq_D15}
\end{equation}
and
\begin{equation}
\pFq{2}{1}\left(\frac{3}{4},\frac{5}{4},\frac{3}{4},z\right)=\frac{1}{(1-z)^{5/4}}.
\label{eq_D16}
\end{equation}
By fixing in Eq.~(\ref{eq_D15}) that $a\!=\!-1/4$, $b\!=\!5/4$ and $c\!=\!3/4$, and
replacing its right-hand side by Eq.~(\ref{eq_D16}), we obtain
\begin{equation}
\frac{\partial}{\partial z}\pFq{2}{1}\left(-\frac{1}{4},\frac{5}{4},\frac{3}{4},z\right)=
\frac{1}{4z}\left[\pFq{2}{1}\left(-\frac{1}{4},\frac{5}{4},\frac{3}{4},z\right)-
\frac{1}{(1-z)^{5/4}}\right].
\label{eq_D17}
\end{equation}
Lastly, upon conducting appropriate calculations in Eq.~(\ref{eq_D14}) with the help of
Eq.~(\ref{eq_D17}) in which $z\!=\!1-t/\tau_{2}(x_{0})$, we achieve the final result:
\begin{equation}
-\frac{\mathrm{d}Q_{2}(t\!\mid\!x_{0})}{\mathrm{d}t}=\frac{\Gamma^{2}\!\left(\frac{1}{4}
\right)\lvert x_{0}\rvert}{8\sqrt{2}\,\pi(Dt)^{1/4}t}\,\Theta\!\left(t-\tau_{2}(x_{0})\right)
\left(\frac{t}{\tau_{2}(x_{0})}-1\right)^{-\frac{1}{4}}.
\label{eq_D19}
\end{equation}

\section{Asymptotic formula for $\boldsymbol{\sigma}$-dependent first-passage time distribution}

A determination of the asymptotic representation for the first-passage time distribution
embodied by Eq.~(\ref{eq_50}) of the main text is rather a difficult endeavor. For this reason,
instead starting with this complex formula embedded in the real space, we will follow another
route of reasoning having the origin in the Laplace space.

To begin with, let us first rewrite the Laplace transform of the survival probability in
Eq.~(\ref{eq_48}) as follows
\begin{equation}
s\,\tilde{Q}_{\sigma}(s\!\mid\!x_{0})\simeq
1-\frac{1}{\Gamma\left(\frac{\sigma+1}{\sigma+2}\right)}
\left[\Gamma\left(\frac{\sigma+1}{\sigma+2},s\eta(x_{0},\sigma)\right)
-\frac{1}{\sigma}\left(s\eta(x_{0},\sigma)\right)^{\frac{2}{\sigma+2}}
\Gamma\left(\frac{\sigma-1}{\sigma+2},s\eta(x_{0},\sigma)\right)\right].
\label{eq_E1}
\end{equation}
By making the Laplace transformation of Eq.~(\ref{eq_11}) complemented with the initial
condition $Q_{\sigma}(0\!\mid\!x_{0})\!=\!1$ for the survival probability, we have
\begin{equation}
\tilde{F}_{\sigma}(s\mid x_{0})=1-s\,\tilde{Q}_{\sigma}(s\!\mid\!x_{0}),
\label{eq_E2}
\end{equation}
while the substitution of Eq.~(\ref{eq_E1}) to the above formula simply gives
\begin{equation}
\tilde{F}_{\sigma}(s\!\mid\!x_{0})\simeq\frac{1}{\Gamma\left(\frac{\sigma+1}{\sigma+2}\right)}
\left[\Gamma\left(\frac{\sigma+1}{\sigma+2},s\eta(x_{0},\sigma)\right)
-\frac{1}{\sigma}\left(s\eta(x_{0},\sigma)\right)^{\frac{2}{\sigma+2}}
\Gamma\left(\frac{\sigma-1}{\sigma+2},s\eta(x_{0},\sigma)\right)\right].
\label{eq_E3}
\end{equation}
In order to learn about the behavior of the first-passage time distribution in the long-time
limit, that is $t\!\to\!\infty$, we need to determine the adequate expression for
Eq.~(\ref{eq_E3}) in the limit $s\!\to\!0$. In this case, it is enough to Taylor expand both
the upper incomplete gamma functions dependent on $s$ in Eq.~(\ref{eq_E3}). Due to the fact 
that $\Gamma(\alpha,z)\simeq\Gamma(\alpha)-\alpha^{-1}z^{\alpha}$ for $z\!\to\!0$, they are
\begin{equation}
\Gamma\left(\frac{\sigma+1}{\sigma+2},s\eta(x_{0},\sigma)\right)\simeq
\Gamma\left(\frac{\sigma+1}{\sigma+2}\right)-\frac{\sigma+2}{\sigma+1}\,
[\eta(x_{0},\sigma)]^{\frac{\sigma+1}{\sigma+2}},
\label{eq_E4}
\end{equation}
and
\begin{equation}
\Gamma\left(\frac{\sigma-1}{\sigma+2},s\eta(x_{0},\sigma)\right)\simeq
\Gamma\left(\frac{\sigma-1}{\sigma+2}\right)-\frac{\sigma+2}{\sigma-1}\,
[\eta(x_{0},\sigma)]^{\frac{\sigma-1}{\sigma+2}}.
\label{eq_E5}
\end{equation}
In this way, Eq.~(\ref{eq_E3}) takes the following form
\begin{equation}
\tilde{F}_{\sigma}(s\!\mid\!x_{0})\simeq1-\left(\frac{\sigma+2}{\sigma+1}-\frac{\sigma+2}
{\sigma(\sigma-1)}\right)\left[s\eta(x_{0},\sigma)\right]^{\frac{\sigma+1}{\sigma+2}}
-\frac{1}{\sigma}\frac{\Gamma\left(\frac{\sigma-1}{\sigma+2}\right)}
{\Gamma\left(\frac{\sigma+1}{\sigma+2}\right)}\left[s\eta(x_{0},\sigma)\right]
^{\frac{2}{\sigma+2}}.
\label{eq_E6}
\end{equation}
In this point we can take advantage on the inverse Laplace transformations
$\mathcal{L}[1;s]\!=\!\delta(t)$ and $\mathcal{L}[(as)^{\alpha};s]\!=\!\frac{1}
{\Gamma(-\alpha)\,t}\left(\frac{a}{t}\right)^{\alpha}$ for $\alpha\!>\!0$, as well as the
product $\Gamma(\alpha)\Gamma(-\alpha)\!=\!-\frac{\pi}{\alpha\sin(\pi\alpha)}$ of the Euler
gamma functions with opposite values of the argument $\alpha$. Thus, performing the inverse
Laplace transformation of Eq.~(\ref{eq_E6}), we readily obtain that
\begin{eqnarray}
F_{\sigma}(t\!\mid\!x_{0})&\propto&\delta(t)+\frac{1}{\pi\sigma t}
\bigg[\frac{\left((\sigma-1)^{2}-2)\right)\sin\left(\frac{\pi}{\sigma+2}\right)}
{\sigma-1}\left(\frac{\eta(x_{0},\sigma)}{t}\right)^{\frac{\sigma+1}{\sigma+2}}
\nonumber\\
&+&\frac{2\sin\left(\frac{2\pi}{\sigma+2}\right)}{\sigma+2}\mathrm{B}\left(\frac{\sigma-1}
{\sigma+2},\frac{2}{\sigma+2}\right)\left(\frac{\eta(x_{0},\sigma)}{t}\right)^{\frac{2}
{\sigma+2}}\bigg],
\label{eq_E7}
\end{eqnarray}
where $\mathrm{B}(\alpha,\beta)\!=\!\frac{\Gamma(\alpha)\Gamma(\beta)}{\Gamma(\alpha+\beta)}$
is the Euler beta function. Considering that $\sigma\!>\!1$ and $t\!\gg\!0$, we can identify
the dominant component in Eq.~(\ref{eq_E7}), which is the expression displayed in the second
line of this formula. Therefore, the asymptotic representation of the $\sigma$-dependent
first-passage time distribution in the first approximation finally reads
\begin{equation}
F_{\sigma}(t\!\mid\!x_{0})\propto\frac{2\sin\left(\frac{2\pi}{\sigma+2}\right)}
{\pi\sigma(\sigma+2)\,t}\,\mathrm{B}\!\left(\frac{\sigma-1}{\sigma+2},\frac{2}{\sigma+2}\right)
\!\left(\frac{\eta(x_{0},\sigma)}{t}\right)^{\frac{2}{\sigma+2}}.
\label{eq_E8}
\end{equation}
%


\bibliography{apssamp}

\end{document}